\documentclass[fleqn,usenatbib]{mnras}

\usepackage{float}
\usepackage{newtxtext,newtxmath}
\usepackage[none]{hyphenat}

\usepackage[T1]{fontenc}
\usepackage{ae,aecompl}
\usepackage{hyperref}
\usepackage{subcaption}
\usepackage{graphicx}	
\usepackage{amsmath}	
\usepackage{amssymb}	
\newcommand{\hi}{{\sc H~i}}
\newcommand{\ovi}{{\sc O~vi}}
\newcommand{\kms}{km s$^{-1}$}

\title[VLT/MUSE galaxy survey towards Q1410]{A VLT/MUSE galaxy survey towards QSO Q1410: looking for a WHIM traced by BLAs in inter-cluster filaments\thanks{Based on observations collected at the European Organization for Astronomical Research in the Southern Hemisphere under ESO programme 094.A-0575(C).}}

\author[I. Pessa et al.]{Ismael Pessa,$^{1}$\thanks{E-mail: ismael.pessa@gmail.com}
Nicolas Tejos,$^{2}$
L. Felipe Barrientos,$^{1,3}$
Jessica Werk,$^{4}$\and
Richard Bielby,$^{5}$
Nelson Padilla,$^{1,6}$
Simon L. Morris,$^{5}$
J. Xavier Prochaska,$^{7}$\and
Sebastian Lopez,$^{8}$
Cameron Hummels$^{9}$
\\
\\
$^{1}$ Instituto de Astrof\'isica, Pontificia Universidad Cat\'olica de Chile, Vicu\~na Mackenna 4860, Santiago, Chile\\
$^{2}$ Instituto de F\'isica, Pontificia Universidad Cat\'olica de Valpara\'iso, Casilla 4059, Valpara\'iso, Chile\\
$^{3}$ Millennium Institute of Astrophysics, Santiago, Chile\\
$^{4}$ Department of Astronomy, University of Washington, Seattle, WA 98195-1580, USA\\
$^{5}$ Department of Physics, Durham University, South Road, Durham, DH1 3LE, United Kingdom\\
$^{6}$ Centro de Astro-Ingenier\'ia, Pontificia Universidad Cat\'olica de Chile, Santiago, Chile\\
$^{7}$ Department of Astronomy and Astrophysics, UCO/Lick Observatory, University of California, 1156 High Street, Santa Cruz, 95064, USA\\
$^{8}$ Departamento de Astronom\'ia, Universidad de Chile, Casilla 36-D, Santiago, Chile\\
$^{9}$ TAPIR, California Institute of Technology, Pasadena, CA 1125, USA\\
}

\date{Accepted XXX. Received YYY; in original form ZZZ}

\pubyear{2018}

\begin{document}
\label{firstpage}
\pagerange{\pageref{firstpage}--\pageref{lastpage}}
\maketitle

\begin{abstract}
Cosmological simulations predict that a significant fraction of the low-$z$ baryon budget resides in large-scale filaments in the form of a diffuse plasma at temperatures $T \sim 10^{5} - 10^{7}$\,K. {However, direct observation of this so-called  warm-hot intergalactic medium (WHIM) has been elusive.} In the $\Lambda$CDM paradigm, galaxy clusters correspond to the nodes of the cosmic web at the intersection of several large-scale filamentary threads. In previous work, we used HST/COS data to conduct the first survey of broad \hi~Ly$\alpha$ absorbers (BLAs) potentially produced by WHIM in inter-cluster filaments. We targeted a single QSO, namely Q1410, whose sight-line intersects $7$ independent inter-cluster axes at impact parameters $<3$\,Mpc {(co-moving)}, and found a tentative excess of a factor of ${\sim}4$ with respect to the field. Here, we further investigate the origin of these BLAs by performing a blind galaxy survey within the Q1410 field using VLT/MUSE. We identified $77$ sources and obtained the redshifts for $52$ of them. Out of the total sample of $7$ BLAs in inter-cluster axes, we found $3$ {\it without} any galaxy counterpart to stringent luminosity limits {($\sim 4 \times 10^{8}$\,L$_{\odot}$ ${\sim} 0.01$\,L$_{*}$)}, providing further evidence that these BLAs may represent genuine WHIM detections. {We combined this sample with other suitable BLAs from the literature and inferred the corresponding baryon mean density for these filaments in the {range 
$\Omega^{\rm fil}_{\rm bar}= 0.02-0.04$}. Our rough estimates are consistent with the predictions from numerical simulations but still subject to large systematic uncertainties, {mostly from the adopted} geometry, ionization corrections and density profile.}
\end{abstract}

\begin{keywords}
{--galaxies: intergalactic medium --techniques: spectroscopic --quasars: absorption lines --methods: observational --cosmology: large-scale structure of Universe}
\end{keywords}



\section{Introduction}
\label{sec:int}
The current cosmological paradigm predicts that only $\sim$4.8$\%$ of the energy content in the Universe is in the form of baryonic matter \citep{Planck2015}. At higher redshifts ($z{\gtrsim}3$) about 90\% of the baryons are assembled in the diffuse photoionized intergalactic medium \citep{Weinberg1997} that give raise to the so called Ly$\alpha$ Forest. In contrast, in the local Universe the fraction of baryons in this phase is only ${\sim}30\%$ \citep[][]{Penton2004, Lehner2007}. From the remaining baryons, only ${\sim}30\%{-}40\%$ are found in other well studied phases (e.g. Stars, ISM, Cluster gas) \citep{Dave2001}, leaving ${\sim}40\%{-}30\%$ of the low-$z$ baryons unaccounted for. This is the so-called `missing baryons problem' {on cosmic scales, where a significant fraction of the total baryons} are missing in the $z{<}1$ Universe \citep{PersicSalucci1992, Fukugita1998, ProchaskaTumlinson2009, Shull2012}. 

Different hydrodynamical cosmological simulations based on $\Lambda$CDM cosmology have predicted that ${\sim}30\%{-}40\%$ of the total baryons at low redshift would be in the warm-hot intergalactic medium (WHIM), at temperatures between $10^5-10^7$\,K particularly residing in diffuse filamentary large-scale structures with a median overdensity of ${\sim}10{-}30$ times the mean density of the Universe \citep{CenOstriker1999, Dave2001}. This is because at the present epoch, hierarchical structure formation model has had time to produce deeper potential wells where baryonic matter is accreted and heated due to the gravitational shocks produced by its collapse. As a consequence of this shock heating, almost all the hydrogen is ionized (by collisional processes or UV radiation) and only a small fraction remains neutral \citep[$f_{\rm HI}{\sim}10^{-5}$ in a pure collisional ionization scenario;][]{SutherlandDopita1993, Richter2004}).

The definitive observational confirmation of the WHIM has been elusive because of the low expected column density of HI in the hot gas ($N_{\rm HI} \approx 10^{13}$\,cm$^{-2}$) \citep{Richter2006b} and large Doppler parameters ({tipically $b\geq 40$\,km s$^{-1}$}; from thermal and non-thermal processes) that would place the absorption features produced by the WHIM at the limit of detectability \citep[e.g.][]{CenOstriker1999,Dave2001}. Emission of this plasma is also expected in the UV and $X$-ray, and marginal detections have been reported \citep[e.g.][]{Hattori2017}; a firm detection in emission still awaits more sensitive telescopes \citep{Fang2005}. More recently, \citet{Tanimura2017} and \citet{Graaff2017} have reported statistically significant detections ($>5\sigma$) of warm-hot baryons through the thermal Sunyaev-Zel'dovich effect signal in a sample of stacked filaments connecting massive haloes. They established a gas density of ${\sim}6$ times the mean universal baryon density, accounting for ${\sim}30\%$ of the total baryon budget. However, absorption line techniques may still represent our current best chance to detect {\it individual} WHIM signatures, particularly through broad \hi~Ly${\alpha}$ absorptions (BLA) in the FUV spectra of bright QSOs.\\

Previous studies have detected BLAs potentially produced by WHIM at low redshifts \citep{Richter2006b, Tripp2006, Lehner2007, Danforth2010,Tilton2012,Wakker2015}. \citet{Richter2006b} calculated the incidence of BLAs per unit redshift to be $dN/dz{\approx}22$ using 4 different QSO sight-lines and derived a lower limit for the baryon content of BLAs $\Omega_{\rm BLA} >0.0027 h_{70}^{-1}$. These results are subject to the uncertainty that not every BLA detected in the FUV QSO spectrum is necessarily related to the WHIM and the authors estimated an associated systematic error as high as $50\%$, that could lead to an overestimation of $\Omega_{\rm BLA}$. These studies assume collisional ionization equilibrium, but according to simulations \citep{FangBryan2001}, photoionization by the UV background also becomes important at typical WHIM densities. Neglecting photoionization can conversely lead to underestimation of the baryon density. 
\citet{Bonamente2016} used {\it Chandra} spectra and found an absorption line identified as O~{\sc VIII} that could potentially be the $X$-ray counterpart of the FUV BLA detected by \citet{Tilton2012}. Indeed, from the SDSS data \citet{Bonamente2016} found evidence of a large-scale filament structure at nearly the same redshift as the absorption features.
{\citet{Wakker2015} used HST spectra of $24$ AGN to sample the gas in a low-$z$ filament, by measuring the properties of $15$ Ly$\alpha$ absorbers in the AGN spectra that are likely associated to the intergalactic gas of the filament. In particular, they studied the properties of the gas as a function of the impact parameter to the filament axis and found evidence that the Ly$\alpha$ line-width anticorrelates with the filament impact parameter. Furthermore, the authors found $4$ BLAs in this sample, all of them in the sight-lines passing relatively close ($<540$\,kpc) to the axis of the filament, which would suggest an increase in temperature and/or turbulence.} 

{In this paper,} we aim at establishing a more accurate relation between {BLAs and the WHIM}. \citet{Tejos2016} performed a novel experiment, searching for BLA features potentially produced by the WHIM at the redshifts where large-scale filaments should exist. They targeted a single QSO at $z\sim 0.79$ (SDSS J141038.39+230447.1; hereafter referred to as Q1410) whose unique sight-line passes throughout $7$ independent cluster pairs at $0.1<z<0.5$, with impact parameters $<3$\,Mpc to the inter-cluster axes connecting them. Theoretical models predict a high probability of finding a filamentary structure between close ($\lesssim 20$\,Mpc) and massive ($\gtrsim 10^{14} \mathrm{M}_{\odot}$) galaxy cluster pairs \citep[e.g.][]{Colberg2005,GonzalezPadilla2010,AragonCalvo2010}. The authors identified $7$ BLAs {with} Doppler parameters $>50$\,km s$^{-1}$ in the spectrum of Q1410 at similar redshifts for $6$ out of the $7$ cluster pairs {and found a tentative excess of BLAs of a factor of ${\sim}4$ with respect to the field}. These BLAs now became potential WHIM signatures and are the subject of further investigation presented in this paper.

We aim at determining the origin of these BLAs, in particular to assess whether these absorption features are produced by the intergalactic medium (IGM) or by the halos of intervening galaxies \citep[e.g.][]{Williams2013}. To discern between these cases, we used the physical impact parameter of nearby galaxies to the Q1410 sight-line and the relative velocity offset between these galaxies and the BLAs.  To this end, we have conducted a blind galaxy survey using the VLT/MUSE IFU \citep{muse2014}, with particular emphasis on the presence or lack of galaxies at the redshifts of the reported BLAs. Our survey used $1$\,hour of VLT/MUSE integration time, reaching redshift completeness level of $\sim75\%$ down to magnitude $r_{\rm AB} = 25$ mag.
\\
Our paper is structured as follows. In Section~\ref{sec:muse_obs} we present and describe our data. In Section~\ref{sec:GalSurvey} we describe our survey, including the identification, characterization, completeness and limitations. Section~\ref{sec:results} presents our main results and in Section~\ref{sec:discussion} we discuss them. A summary of our results and conclusions are presented in Section~\ref{sec:conc}. For our analysis, we assume a $\Lambda$CDM cosmology based on the results of the \citet{Planck2015}.

\section{Data}
\label{sec:muse_obs} 
\subsection{VLT/MUSE Integral field spectroscopy}
We obtained VLT/MUSE data of a $\sim 1 \times 1$\,arcmin$^2$ field containing Q1410 as part of the ESO programme 094.A-0575 (PI Tejos). The observations were taken with a seeing of $\sim$0.8", sampled at $0.2\times0.2$\,arcsec$^2$, with a spectral range from $4750-9350$\,\AA, and a resolving power $ R\sim 1770-3590$. A total of $4$ exposures of $15$\,min each were used, $2$ of them centered on the Q1410 field with a position angle PA$=0^{\circ}$ and the remaining $2$ centered on the brightest nearby galaxy at RA$=$14h10m39.8s and Dec.$=$+23d05m00.8s (J2000; $\Delta\mathrm{RA}{\approx} 19.4$", $\Delta\mathrm{Dec.}{\approx}14.2$") with a PA${=}90^{\circ}$. As a result of this double pointing we obtained higher signal-to-noise data where the $2$ fields overlap for a full $1$\,hr exposure. In the edges, the effective exposure time is $30$\,min. Table~\ref{tab:point} summarizes these observations and Fig.~\ref{fig:musefov} shows the targeted MUSE field (grey contours). \begin{table}{\large VLT/MUSE observations}
\centering
\label{tab:}
\begin{tabular}{lccr}
\hline
\multicolumn{2}{c}{Pointings}&Exposure time&PA\\
          RA&DEC & s&deg\\
\hline
14:10:38.39&+23:04:47.18&$2 \times 900$&0\\
14:10:39.79&+23:05:00.80&$2 \times 900$&90\\
\hline
\end{tabular}
\caption{\label{tab:point}Summary of our observations described in Sect. 2. }
\end{table}
Datacubes have been reduced and combined using the standard MUSE pipeline version muse-1.4 (see \url{http://www.eso.org/observing/dfo/quality/PHOENIX/MUSE/processing.html} for details). As a post-processing reduction, we used the ZAP software \citep{Soto2016} to perform a second order sky subtraction, which uses principal components analysis (PCA) to isolate and remove residual sky subtraction features.
\label{sec:muse_redux} 

\subsection{VLT/VIMOS Imaging}

Additional VLT/VIMOS data are available for this field. In particular, $2$\,min $R$-band pre-imaging was obtained as part of ESO programme 094.A-0575 (PI Tejos) for the mask preparation of multi-object spectroscopy (MOS) on the field. Unfortunately, the MOS data were never obtained. However, we are able to use the pre-imaging data as an astrometry reference frame for our MUSE data (see Fig.~\ref{fig:musefov}).


\section{Galaxy survey}
\label{sec:GalSurvey}

\subsection{Source identification}
\label{sec:muse_source}

To identify the sources in the field and determine the aperture that best defines the spaxels\footnote{We refer to spaxel as a spatial sampling element, that correspond to a one-pixel spectrum in the datacube.} within each source we used {\sc SExtractor} \citep{Sextractor1996} in the `white image'. This image corresponds to the sum of the flux at all wavelengths for each spaxel.
\\
After performing a visual inspection to remove artifacts, we manually added $10$ sources that were not detected by {\sc SExtractor} but were still visible (but faint) in the `white image'.
\\
Finally, we used MUSELET (for MUSE Line Emission Tracker) a {\sc SExtractor}-based python tool to detect emission lines in a datacube included in the MPDAF Python package \citep{MPDAF} to search for additional emission-lines-only galaxies that were not visible in the `white image' but show at least one strong emission line, and we added $1$ source. \\
We ended up with a sample of $77$ sources in the MUSE FoV, including the central QSO. For each source we extracted a 1-D spectrum combining the spaxels inside the source aperture. The flux on each spaxel was weighted by a `bright profile', proportional to the total flux of that spaxel in the white image. For the sources that were not originally detected by {\sc SExtractor}, the aperture was arbitrarily defined as a circle containing most of the apparent flux. We estimated the $r_{\rm AB}$ magnitude for all the sources by convolving the datacube fluxes with a SDSS $r$ transmission curve. This effectively creates a MUSE $r$-image. The zero-point of this image was calibrated by doing a linear fit with slope of unity between the SDSS {modeled magnitudes} and the {\sc SExtractor} {MAG AUTO} magnitudes\footnote{{We expect that the flux lost by the aperture defined by {\sc SExtractor} accounts for less than $0.1$\,mag.}} of several cross-matched sources in the field (the photometry was only computed for the sources detected by {\sc SExtractor}). Our identified sources are summarized in Table~\ref{tab:ids}.

\subsection{Source characterization}
\label{sec:muse_source2}

Redshifts of each source were measured using Redmonster \citep{Hutchinson2016}. The code performs a $\chi^2$  minimization between the observed spectrum and a set of theoretical models for galaxies, stars and QSOs. These templates are modulated by a low-order polynomial mimicking the effects of galactic extinction, sky-subtraction residual and possible spectrophotometric errors. We adopted a reliability scheme for the redshifts measurements as follows:

\begin{itemize}
\item `a' sources: these are the best characterized, showing at least $2$ well characterized features in their spectra.
\item `b' sources: these are relatively well characterized, showing at least $1$ well identified feature and a possible second feature (marginal) in their spectra. 
\item `c' sources: these are uncertain, showing only one feature in their spectra (typically a single emission line). We included 'c' sources in our analysis in order to be conservative for ruling out the presence of galaxies near BLA features.
\item `d' sources: these could not be characterized as they did not show any spectral feature. Most of them are fainter than $r_{\rm AB}=24$\,mag, and were excluded from our analysis.
\end{itemize}

Out of the $77$ sources, we successfully obtained the redshifts for $52$ of them (see Table~\ref{tab:ids}). These {include $42$ obtained with Redmonster and $10$ from visual inspection of strong emission lines.} In Appendix~\ref{sec:appims} we present the Redmonster fits for the galaxies with a reliability level of `a', `b' and `c'. {For the galaxies showing} only one strong emission line, we assume it to be H$\alpha$, or [OII] if the emission line show a double peak profile. In these cases, the redshift was determined by fitting a Gaussian profile to the emission feature (a double Gaussian profile with a fixed separation was used for the [OII] doublet) and we report only the redshift solution (without the uncertainties). The remaining $25$ that could not be identified are listed in Table~\ref{tab:noids} in the Appendix~\ref{sec:noidsources}.\\

{We empirically estimated the uncertainties for the redshifts obtained with Redmonster \citep{Hutchinson2016} by comparing the redshifts obtained for the same source in each individual exposure (see Section~\ref{sec:muse_obs}). We used a sub-sample of $25$ sources for which Redmonster converged to a fit in at least $2$ individual exposures (out of a total of $42$ sources characterized with redshifts in the combined datacube), and studied the distribution of the differences between the redshifts obtained for each individual source. The dispersion in this distribution is due to the uncertainties of the redshift measurement in both exposures, and we determined its standard deviation to be $\sigma_{\rm diff}{\approx}0.0002$. Therefore, assuming that both exposures have the same uncertainty, we can estimate an individual $\sigma_{z}{\approx}\frac{\sigma_{\rm diff}}{\sqrt[]{2}}{\approx}0.00014$ for the Redmonster redshift measurements. Additionally, we looked for a cross-match between the SDSS spectroscopic catalog and our survey. We found $1$ single source (the central QSO) to match, and the redshift difference was consistent with our estimated $\sigma_{z}$.}

We classified sources with redshifts based on their spectral types as follows:
\begin{itemize}
\item Star-forming (\emph{SF}): galaxies that show strong emission lines and a blue continuum. These 38 sources correspond to $73\%$ of the sample with measured redshifts.
\item Non-star-forming (\emph{non-SF}): galaxies that show a strong red continuum and an absence of emission lines. These  6 sources correspond to ${\sim}11\%$ of the sample with measured redshifts.
\item Red star-forming (\emph{SF-red}): These galaxies show both a strong red continuum and emission lines consistent with recent star formation events. These sources are rare accounting for ${\sim}6\%$ of the sample with measured redshifts.
\item QSO: We find a quasar-like spectrum for 3 sources, corresponding to ${\sim}6\%$ of the sample.
\item Star: stellar-like spectrum. We identified a single star in our field. 
\item Ly$\alpha$ emitter candidate (LAE): One source showed a prominent single emission line, asymmetric and extended to the red, at a $\lambda \approx 5045$ \AA. Given its observed wavelength,  this can not be explained as an H$\alpha$ emission. Furthermore, its profile and strength  are inconsistent with this line being [OII]. Thus, we deem this source as a Ly$\alpha$ emitter candidate at $z{\sim} 3.15$.

\end{itemize}

\begin{figure*}

	\includegraphics[width=0.95\textwidth]{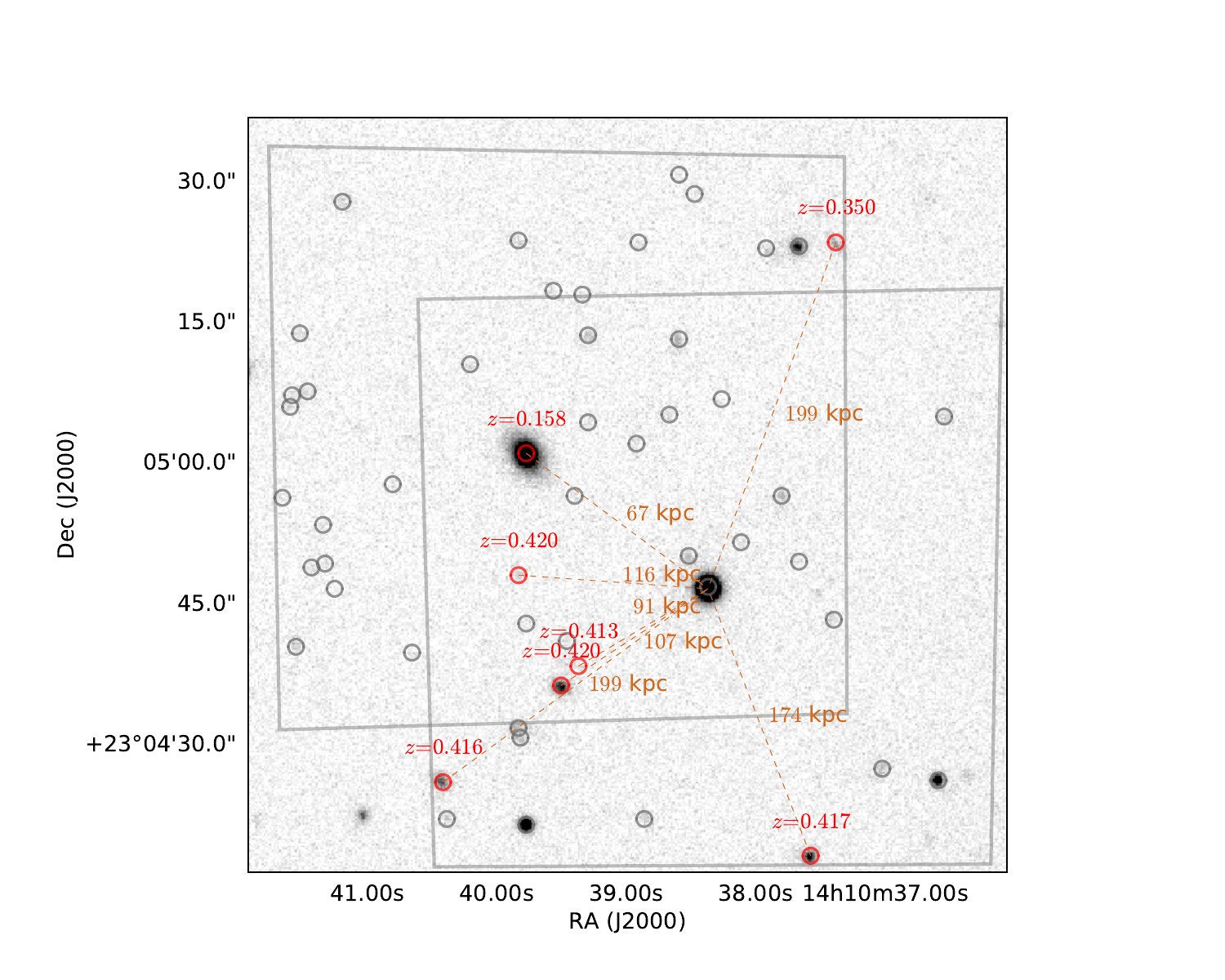}
    \caption{\label{fig:musefov}VLT/VIMOS image of our full MUSE field of view. The VLT/MUSE exposures are centered on the bright galaxy (upper left of the center) and the QSO Q1410 (lower right of the center) respectively (see Table~\ref{tab:point}). The gray contours mark the field of view of each individual VLT/MUSE exposure. We have characterized and measured redshifts for most sources brighter than $r=25$ mag (see Table~\ref{tab:ids})). Red circles show galaxies at redshifts within $\Delta v$ = $\pm$1000 km s$^{-1}$ of any BLA. Grey circles mark the rest of the  sources characterized with a redshift.}
\end{figure*}

The distribution of sources in the field is shown in Fig.~\ref{fig:musefov}. For each source with a redshift we estimated the proper transverse distance to the QSO sight-line using our adopted Planck 2016 cosmology. We also calculated their absolute $r_{\rm AB}$ magnitude as follows:

\begin{equation}
M_{\rm r} = r_{\rm AB} - 5 (\mathrm{log}(d_{\rm L})-1) - \mathrm{K}_{\rm corr}
\end{equation}

\noindent where $d_{\rm L}$ is the luminosity distance in parsecs. and K$_{\rm corr}$ is the corresponding K-correction for each galaxy (see Appendix~\ref{sec:kcorr}).

\subsection{Survey characterization}
\label{sec:muse_survey}

\begin{figure*}
        \begin{subfigure}[b]{0.33\textwidth}
                \includegraphics[width=\columnwidth]{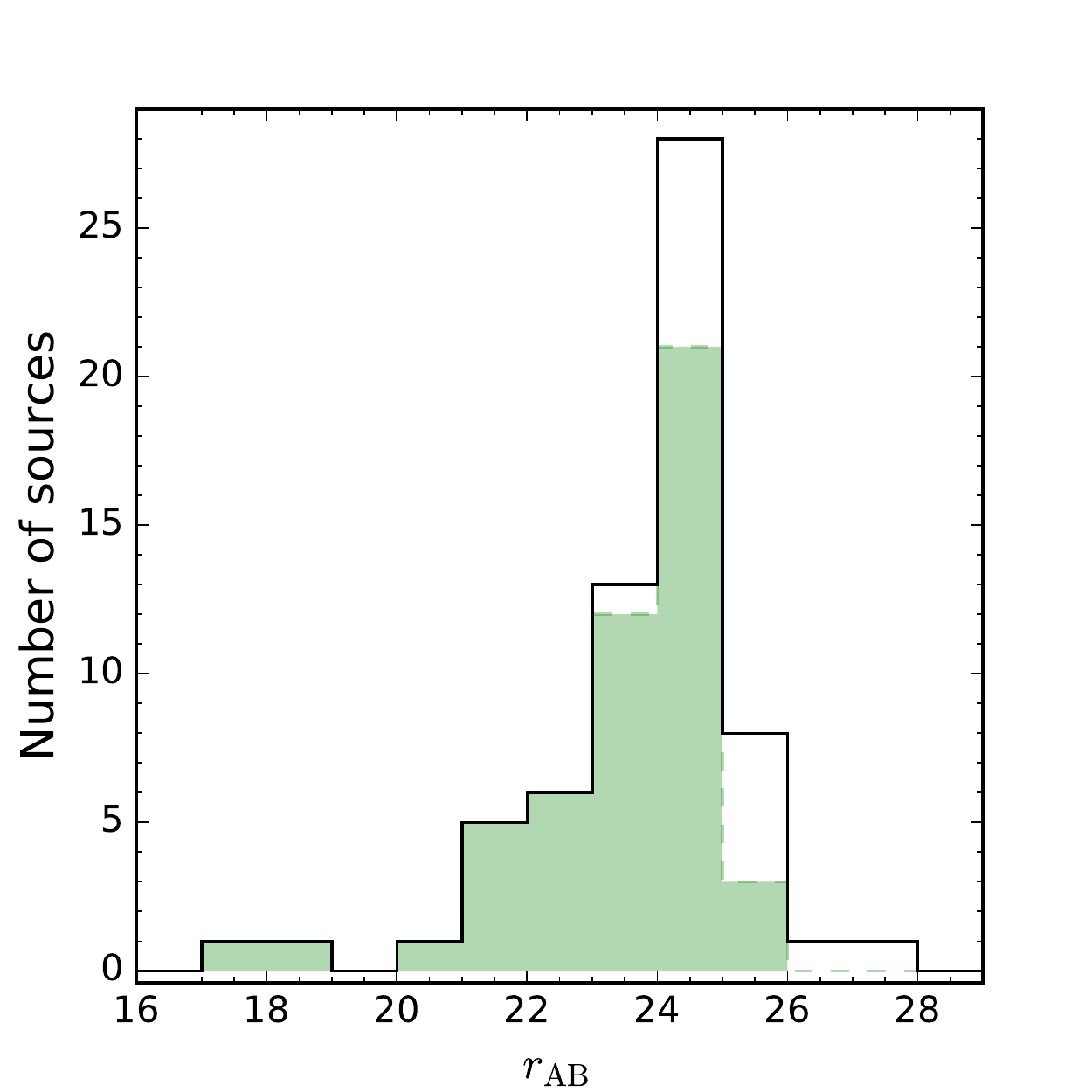}

        \end{subfigure}%
        \begin{subfigure}[b]{0.33\textwidth}
                \includegraphics[width=\columnwidth]{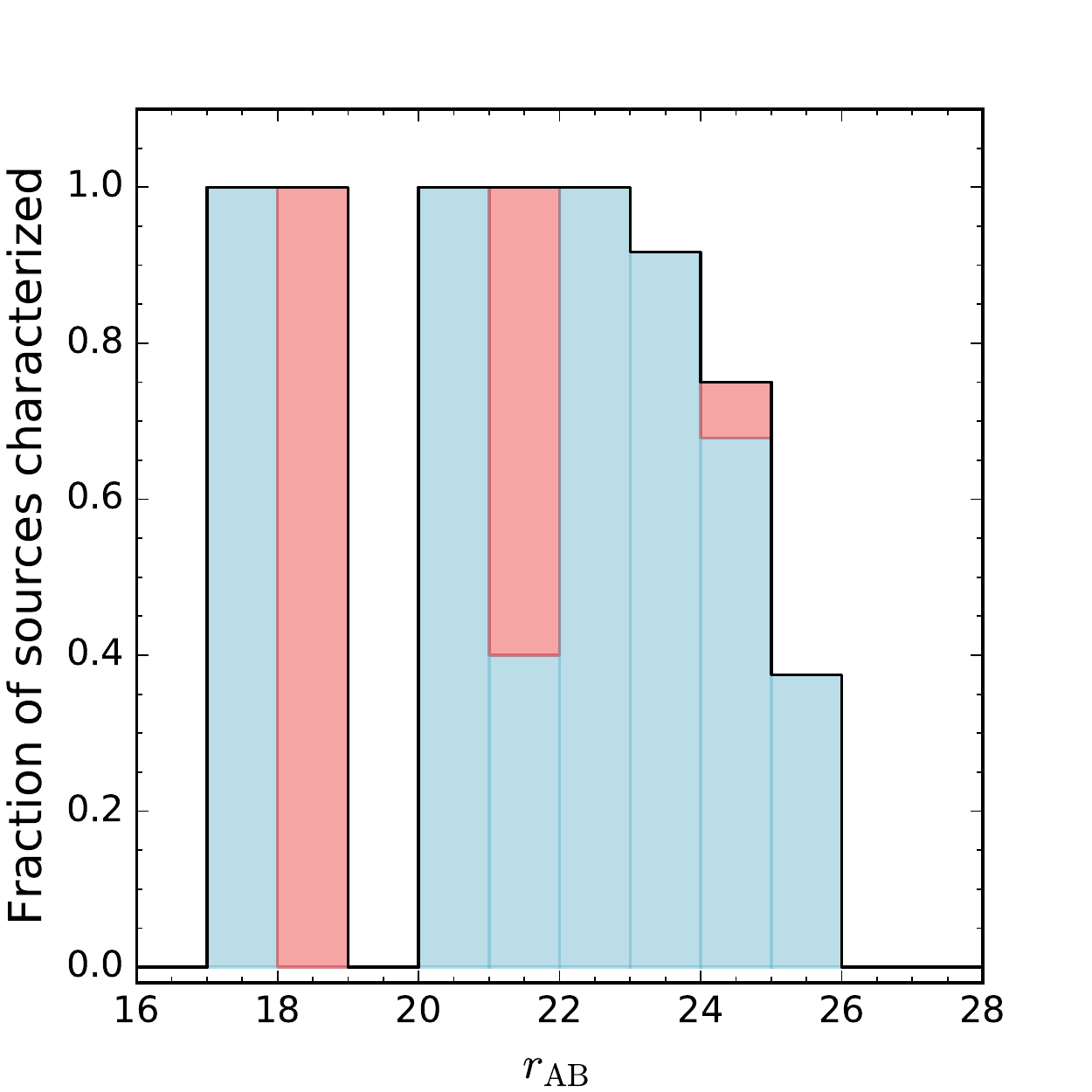}

        \end{subfigure}%
        \begin{subfigure}[b]{0.33\textwidth}
                \includegraphics[width=\columnwidth]{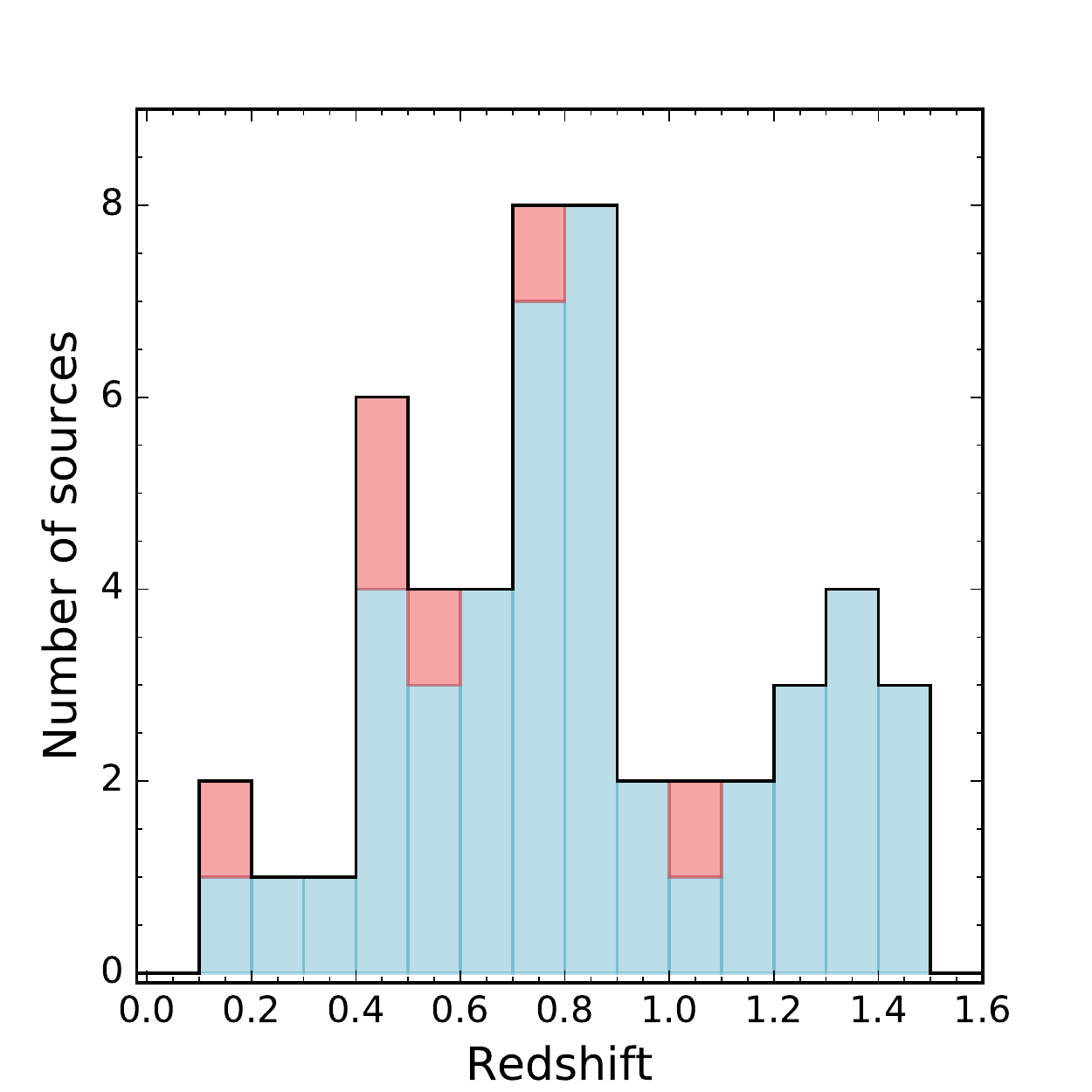}

        \end{subfigure}%

        \caption{\label{fig:hists} \textit{Left}: Survey histogram colored in green for our sample with measured redshifts. The black line shows the distribution for the whole sample detected by SExtractor. Our detection threshold is at $r\sim25$ mag. \textit{Center} shows the completeness fraction of the redshift survey. The star forming galaxies are shown in blue and the non-star forming galaxies in red. The black contour marks our full sample. We reach to $\sim75\%$ successful characterization fraction at an apparent magnitude $r_{\rm AB}=25$ mag. \textit{Right}: shows the redshift distribution of our full characterized sample colored by star forming and non-star forming fraction same as the central panel. The LAE candidate at $z\approx3.15$ is not shown here.}
\end{figure*}

In order to characterize the completeness of our survey we have used the apparent $r_{\rm AB}$ magnitudes. The left panel of Fig.~\ref{fig:hists} shows a histogram of sources per apparent $r_{\rm AB}$ magnitude bin,  separating the sample for which good redshifts were obtained from the full sample. Our survey peaks at $r_{\rm AB} \approx 25$\,mag. The sudden decline in the number of sources to fainter magnitudes marks our completeness limit. An apparent magnitude of $r_{\rm AB}=25$\,mag suggests a luminosity limit of $\sim 5 \times \ 10^{8}$\,L$_{\odot}$ at $z \sim 0.45$. The central panel of the Fig.~\ref{fig:hists} shows the fraction of sources that were successfully assigned a redshift as a function of $r_{\rm AB}$ bin. Our characterization reaches $\sim 75\%$ for $r_{\rm AB}\approx 25$\,mag. The right panel shows the redshift distribution of our full sample colored by spectral type.
\\

\begin{figure*}
	\includegraphics[width=1.\textwidth]{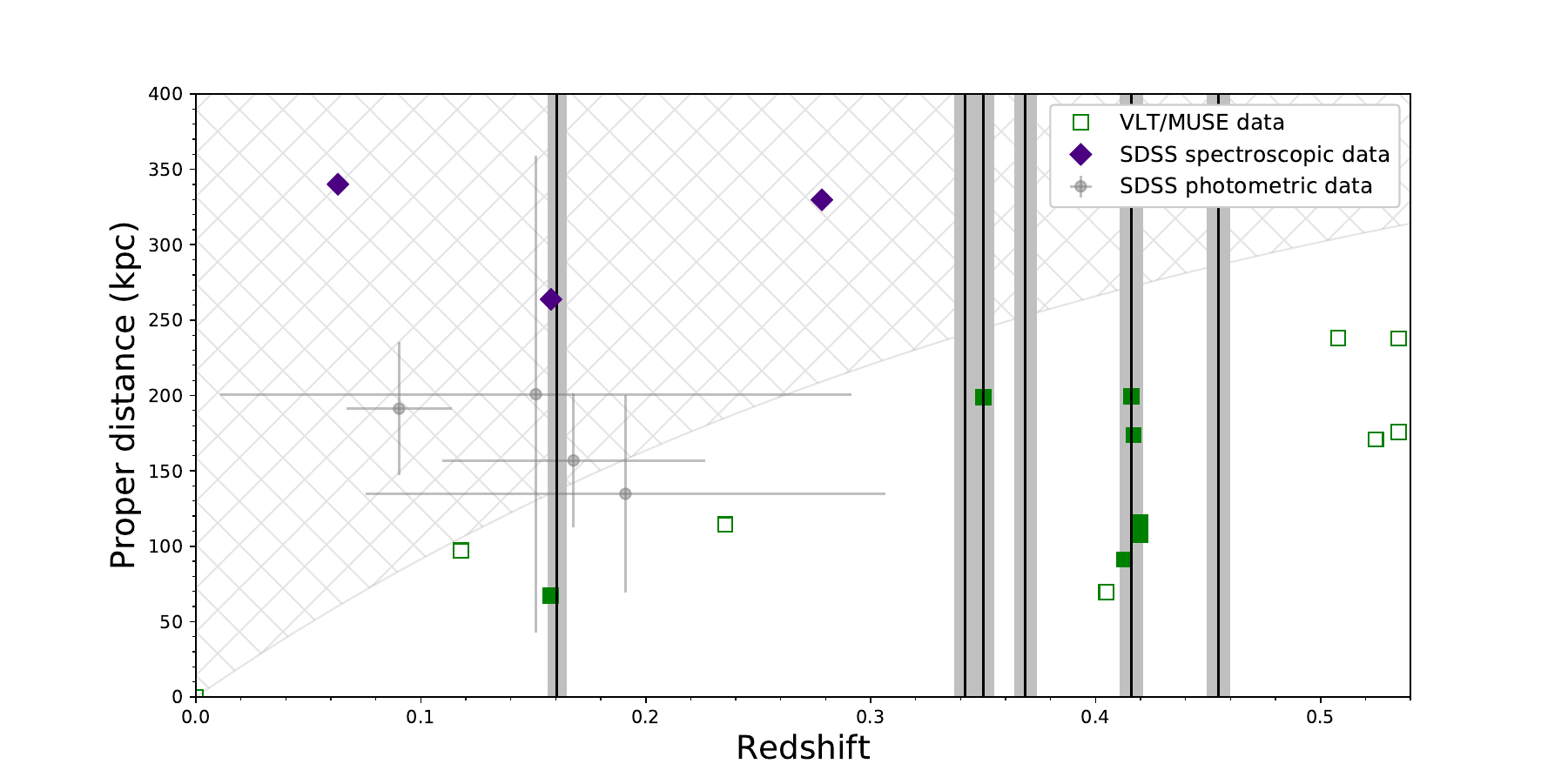}
    \caption{\label{fig:dist}Distribution of impact parameter to the Q1410 sight-line as a function of redshift. Our VLT/MUSE sample of galaxies is shown in green squares. SDSS galaxies are shown in violet diamonds {and gray circles (See Sect.~\ref{sec:sdss_gal} for details). We only show here galaxies in the SDSS photometric catalog that are outside the MUSE FoV.} The vertical black lines surrounded by a shaded region mark the redshift of the inter-cluster axes $\pm1000$ km s$^{-1}$ from its rest frame. Filled squares mark sources in the VLT/MUSE sample that are located within $\pm$1000 km s$^{-1}$ of an inter-cluster axis and will be the subject of further analysis. {The hatched area in the upper left corner is outside the MUSE FoV and shows the approximate edges of our spectroscopic survey.} }
\end{figure*}

Fig.~\ref{fig:dist} shows the distribution of impact parameter to the Q1410 sight-line as a function of redshift. The hatched area in the upper left corner shows the limit of the FoV of VLT/MUSE as a function of redshift; regions in the hatched area are out of the effective VLT/MUSE coverage. At $z\approx 0.1$ our effective radial coverage is about $100$\,kpc, while at $z\approx 0.5$ we reach scales $> 300$\,kpc. The vertical black lines mark the redshift of the inter-cluster axes reported by \citet{Tejos2016} that show a BLA. The vertical dashed regions around these lines represent a rest-frame velocity window of $\Delta v = \pm 1000$\,km s$^{-1}$ around these redshifts. Excluding galaxies at the redshifts of the cluster-pairs, we found $9$ sources brighter than apparent magnitude $r_{\rm AB} = 23$\,mag. This represents a density of $22\,500 \pm 7\,500$\,deg$^{-2}$, assuming a Poissonian error, which is consistent with the density of $ \sim 20\,000$\,deg$^{-2}$ found by the VVDS survey \citep{LeFevre2005}.

Table~\ref{tab:ids} summarizes the characterization of the sources in our full survey.

\newpage
\begin{table*}{\large Sources characterized in our survey}
\centering
\begin{tabular}{lccccrcccll}
\hline
ID&Object&RA&DEC&\multicolumn{2}{c}{Impact Parameter}&$r_{\rm AB}$&$M_{r}$&$z$&class&reliability\\
          & & J2000 &J2000   &    (arcsecs)&(kpc)                      &       &       & &     &\\
(1)  &(2)&(3)&  (4)   &  (5)                      & (6)   &  (7)  &(8)&(9)&(10)&(11)\\
\hline
1&J141038.37+230446.5&212.65987&23.07958&0.0&0&17.35&-26.20&0.7961&QSO&a\\
2&J141038.52+230449.7&212.66050&23.08047&3.8&29&24.15&-22.08&0.7976&SF-red&a\\
3&J141038.11+230451.3&212.65879&23.08092&6.0&46&25.21&-19.31&0.7966&SF&c\\
4&J141037.66+230449.2&212.65692&23.08033&10.2&79&24.39&-18.75&0.8315&SF&a\\
5&J141037.80+230456.2&212.65750&23.08228&12.5&70&23.55&-19.20&0.4048&SF&b\\
6&J141037.40+230443.0&212.65583&23.07861&13.8&107&24.48&-19.91&0.7940&SF&b\\
7&J141039.46+230440.7&212.66442&23.07797&16.1&112&23.35&-19.55&0.6171&SF&a\\
8*&J141039.37+230438.1&212.66404&23.07725&16.2&91&24.77&-17.16&0.4126&SF&c\\
9&J141038.92+230501.8&212.66217&23.08383&17.1&137&24.82&-21.22&0.9093&SF&b\\
10&J141039.40+230456.2&212.66417&23.08228&17.2&149&24.94&-20.02&1.4865&SF&b$^{\bot}$\\
11&J141038.67+230504.8&212.66112&23.08467&18.8&136&24.37&-19.41&0.6750&SF&a\\
12*&J141039.50+230435.9&212.66458&23.07664&18.9&107&21.32&-21.29&0.4199&SF-red&a\\
13&J141039.77+230442.5&212.66571&23.07847&19.7&171&24.29&-19.55&1.4596&SF&b$^{\bot}$\\
14&J141038.27+230506.4&212.65946&23.08511&19.9&144&-&-&0.6676&SF&c\\
15*&J141039.84+230447.8&212.66600&23.07994&20.3&116&23.71&-18.83&0.4198&SF&a\\
16&J141039.29+230504.1&212.66371&23.08447&21.7&169&24.31&-146.55&3.1500&LAE&c$^{\bot}$\\
17*&J141039.77+230500.7&212.66571&23.08353&24.0&67&18.93&-20.94&0.1577&non-SF&a\\
18&J141039.83+230431.4&212.66596&23.07539&25.2&195&22.10&-22.41&0.8125&SF&a\\
19&J141038.86+230421.8&212.66192&23.07272&25.6&197&23.24&-21.14&0.7963&SF&b\\
20&J141039.82+230430.4&212.66592&23.07511&25.7&219&23.12&-20.73&1.2070&SF&b$^{\bot}$\\
21&J141038.60+230512.8&212.66083&23.08689&26.5&171&22.78&-20.43&0.5246&SF&a\\
22&J141037.02+230427.0&212.65425&23.07417&27.0&176&22.90&-19.99&0.5348&SF&a\\
23&J141039.29+230513.3&212.66371&23.08703&29.7&114&23.07&-17.71&0.2354&SF&a\\
24*&J141037.58+230417.9&212.65658&23.07164&30.6&174&21.96&-21.12&0.4169&non-SF&a\\
25&J141036.54+230504.7&212.65225&23.08464&31.1&269&23.92&-22.54&1.3583&SF&b$^{\bot}$\\
26&J141039.78+230421.1&212.66575&23.07253&32.0&256&20.44&-23.43&0.8975&QSO&b\\
27&J141036.59+230425.9&212.65246&23.07386&32.1&277&21.13&-23.82&1.3400&QSO&b\\
28&J141040.65+230439.5&212.66937&23.07764&32.2&278&25.69&-17.36&1.3118&SF&b$^{\bot}$\\
29&J141039.34+230517.6&212.66392&23.08822&33.9&289&23.69&-21.48&1.2081&SF&c$^{\bot}$\\
30&J141040.20+230510.2&212.66750&23.08617&34.6&300&24.32&-21.34&1.4305&SF&b$^{\bot}$\\
31*&J141040.42+230425.6&212.66842&23.07378&35.2&199&21.51&-21.26&0.4159&non-SF&a\\
32&J141040.81+230457.3&212.67004&23.08258&35.4&289&99.00&53.56&0.9722&SF&b\\
33&J141039.56+230518.1&212.66483&23.08836&35.6&308&24.44&-22.46&1.3849&SF&b$^{\bot}$\\
34&J141037.92+230522.5&212.65800&23.08958&36.5&238&22.41&-20.53&0.5347&SF&a\\
35&J141040.38+230421.7&212.66825&23.07269&37.2&-&23.40&-&0.0000&Star&a\\
36&J141038.91+230523.1&212.66212&23.08975&37.4&318&24.71&-21.80&1.1972&SF&b\\
37&J141037.67+230522.8&212.65696&23.08967&37.6&238&21.44&-22.41&0.5079&non-SF&a\\
38*&J141037.38+230523.1&212.65575&23.08975&39.1&199&22.39&-19.69&0.3502&SF&a\\
39&J141041.26+230446.3&212.67192&23.07953&39.9&307&24.94&-27.97&0.7898&non-SF&c\\
40&J141041.33+230448.9&212.67221&23.08025&40.9&327&24.78&-21.48&0.8960&SF&a\\
41&J141041.34+230453.0&212.67225&23.08139&41.5&332&24.39&-27.77&0.8964&SF&b\\
42&J141038.48+230528.4&212.66033&23.09122&41.9&325&24.95&-21.32&0.8123&SF&c\\
43&J141039.83+230523.3&212.66596&23.08981&42.0&297&23.00&-20.95&0.6369&SF&a\\
44&J141041.44+230448.6&212.67267&23.08017&42.4&340&25.09&-31.85&0.8979&SF-red&a\\
45&J141038.59+230530.5&212.66079&23.09181&44.1&97&24.47&-14.58&0.1180&SF&c\\
46&J141041.55+230440.1&212.67312&23.07781&44.3&339&23.35&-22.45&0.7741&SF&a\\
47&J141041.66+230455.9&212.67358&23.08219&46.4&370&24.52&-31.97&0.8889&SF&b\\
48&J141041.47+230507.3&212.67279&23.08536&47.6&406&24.01&-22.36&1.2077&SF&b$^{\bot}$\\
49&J141041.59+230505.7&212.67329&23.08492&48.4&358&23.84&-21.85&0.7086&SF&a\\
50&J141041.58+230506.9&212.67325&23.08525&48.8&408&24.90&-24.02&1.0799&SF&b\\
51&J141041.52+230513.5&212.67300&23.08708&51.2&428&24.71&-39.93&1.0793&non-SF&a\\
52&J141041.19+230527.5&212.67162&23.09097&56.5&481&23.13&-22.36&1.1980&SF&c\\
\hline
\end{tabular}
\caption{\label{tab:ids}($^{\bot}$): Redmonster did not converge to a $z$ on these sources. Redshift were calculated by a visual inspection on these cases.\\(*): Nearby Galaxies to the inter-cluster filaments. These sources are marked in red in Fig.~\ref{fig:musefov}.\\  Source were classificated according to their spectral type. SF galaxies show strong emission lines and a blue continuum, non-SF galaxies show a strong red continuum and an absence of emission lines and SF-red galaxies show a strong red continuum and emission lines. We also identified a Ly$\alpha$ emitter candidate, which is classificated as LAE. Sources where $r$ is undefined were not detected by SExtractor and we manually included them in this survey. {The uncertainties of the Redmonster redshift measurements in the Col. (9) are of the order of ${\sim}$0.00014}.}
\end{table*}

\subsection{SDSS galaxies}
\label{sec:sdss_gal}
As mentioned before, at low-$z$ our VLT/MUSE survey has a relatively limited FoV. In particular, at $z \approx 0.16$ we would need to triplicate our current FoV to cover a physical impact parameter of $ \sim 300$\,kpc, which has been suggested for the so-called circum-galactic medium \citep[CGM;][]{Prochaska2011, Borthakur2016}. In order to partially compensate this limitation, we have included galaxies with spectroscopic redshifts from the SDSS \citep{SDSS13}. These galaxies are marked with diamonds in Fig.~\ref{fig:dist}. At $z \approx 0.16$ we found $3$ SDSS galaxies, the closest being at an impact parameter of $\sim 260$\,kpc to the QSO sightline. The remaining $2$ are located at an impact parameter of ${>}450$\,kpc and are not shown in Fig.~\ref{fig:dist} (See Table ~\ref{tab:SDSS}).
\\
{We have also looked for evidence of galaxies that we could be missing outside the MUSE FoV at the redshifts of the inter-cluster axis in the SDSS photometric catalog. We found ${\sim}400$ additional galaxies closer than $4$\arcmin\ around the line of sight, corresponding to a transverse distance of ${\sim}500$\,kpc at $z=0.1$. Given the high uncertainties in the photometric redshifts (and consequently, in the estimated physical impact parameters), we have looked only for galaxies whose physical impact parameters may lie within $2$ times their inferred virial radii to the QSO sightline, given their inferred stellar masses. Stellar masses were obtained from their photometry using the relation presented in \citet{Taylor2011}. We then estimated their virial masses and virial radii in the same fashion as explained in Section~\ref{sec:GalAtBLAz} below. To be conservative, we considered the lower value of the error-bar for the physical impact parameter and the upper value of the error-bar for the virial radius (this uncertainty comes from the dispersion of ${\sim}$0.1\,dex in the stellar mass relation of \citet{Taylor2011}). We found $8$ sources satisfying this condition between its impact parameter and virial radii at photometric redshifts $z<0.5$, of which $4$ are outside the MUSE FoV and are shown in Fig.~\ref{fig:dist} as gray circles (the other $4$ were already detected by our MUSE spectroscopic survey). We found that $2$ of these have photometric redshifts $z\lesssim 0.2$ with relatively low redshift uncertainties, making them unlikely to be associated to any of our BLAs at $z>0.3$. However, the remaining 2 need further analysis.
Given that the uncertainties in the inferred proper impact parameter comes from the large uncertainty in the photometric redshift estimation, we can check if these galaxies could be related to the BLA at $z=0.3422$ (or higher), by comparing their virial radius with the impact parameter that would have at that redshift. If we set these galaxies to be at that redshift, we obtain that their impact parameter would be higher than $4$ times their estimated virial radius; if one of those galaxies is at z${\approx}0.3422$ (or higher), it would be at a large enough impact parameter for it not to be physically associated with the BLA. Thus, we conclude that we are likely not missing any galaxy outside the FoV of MUSE that could be directly associated to a BLA at redshift $z=0.3422$ (or higher).}

\section{Results}
\label{sec:results}

\subsection{Galaxies at BLA redshifts}
\label{sec:GalAtBLAz}

We now focus on galaxies that may be related to the BLAs of interest, i.e. those observed at redshifts of cluster-pairs as reported by \citet{Tejos2016}. From our blind galaxy survey, we selected the subsample of galaxies lying within $\Delta v \pm 1000$\,km s$^{-1}$ from any of the aforementioned BLAs. {We found 7 out of the 52 characterized sources} of the sample satisfying this criterion (filled green squares in Fig.~\ref{fig:dist}), for which stellar masses, halo masses, virial radii and virial velocity dispersions were estimated. Stellar masses were calculated using the StarLight software \citep{CidFernandez2005}, which performs a spectral synthesis analysis assuming a \citet{Chabrier2003} initial mass function (IMF). Since StarLight does not provide an estimation of the uncertainty for the inferred stellar masses, we assume its error to be $0.25$\,dex based on the analysis presented in \citet[][see their figure 1]{Li2017}. Halo masses were estimated from the stellar masses by assuming the bijective relation between the two as given by \citet{Moster2010}. We consider this inferred halo mass as a virial mass, $M_{\rm vir}$ of a galactic system and estimated a virial radius as, 

\begin{equation}
R_{200} = \bigg(\frac{M_{\rm vir}}{\frac{4}{3}\pi200\rho_{\rm c}(z)}\bigg)^{1/3}
\end{equation}

\noindent i.e., $R_{200}$ is the radius of the spherical volume where $M_{\rm vir}$ is contained at $200$ times the critical density of the Universe at a given redshift, $\rho_{\rm c}(z)$. Velocity dispersions for each galaxy were estimated as,

\begin{equation}
\sigma_{\rm vir} = \sqrt[]{\frac{GM_{\rm vir}}{R_{200}}}
\end{equation}

\noindent Uncertainties in $M_{\rm vir}$, $R_{200}$, and $\sigma_{\rm vir}$ where estimated by propagating the adopted error in the stellar masses. Table~\ref{tab:gals} summarizes these inferred properties for our sample galaxies near BLAs.

In the following, we use the scales given by $R_{200}$ and $\sigma_{\rm vir}$ of each galaxy to discriminate whether it could be directly associated to a BLA feature or not.  

\subsection{Potential association of galaxies to known BLAs}
\label{sec:asso}

In \citet{Tejos2016} the authors identified $7$ H~{\sc i} broad Ly${\alpha}$ absorption features at $6$ different redshifts, related to $6$ out of the $7$ inter-cluster axes at $0.1<z<0.5$. As seen in Fig.~\ref{fig:dist}, $3$ out of the $6$ relevant inter-cluster axes show at least one nearby galaxy.\footnote{Note that there are $2$ BLAs at $z \approx 0.416$, for which there is also a group of galaxies (see Table~\ref{tab:BLAs}).}

\begin{figure}
    \begin{subfigure}[t]{0.48\textwidth}
	\includegraphics[width=1.15\columnwidth]{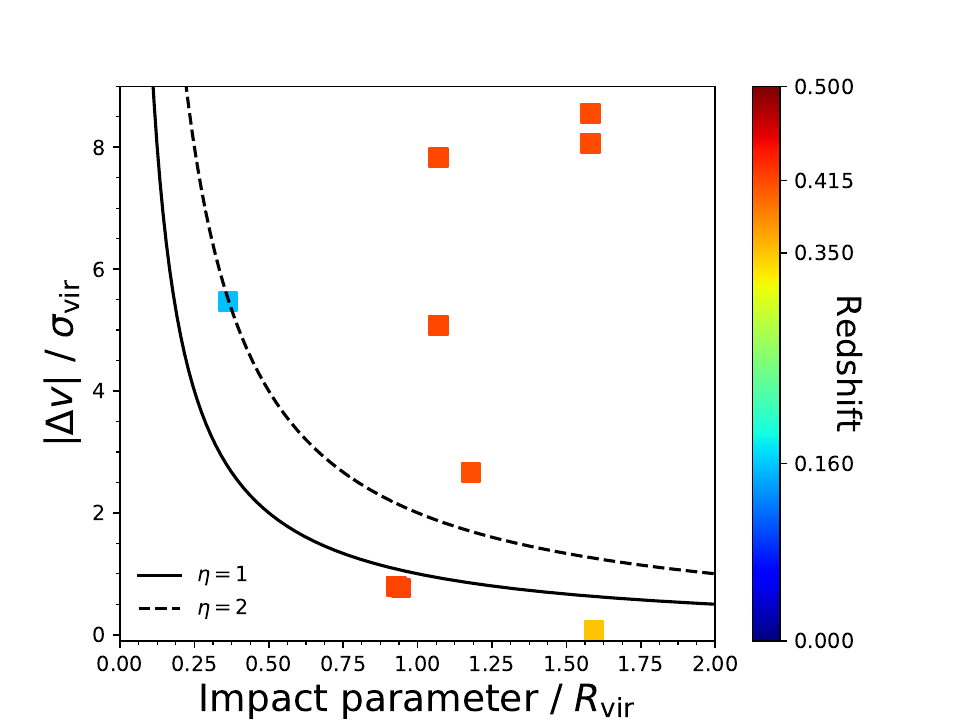}
    \end{subfigure}
    \caption{\label{fig:blas}Each square represents a single galaxy-BLA combination identified from our sample, representing a total of 7 galaxies and 4 absorbers across three separate intra-cluster axes. The  panel shows the relationship between potential host galaxies and absorbers in terms of a velocity difference and impact parameter to the QSO sight-line in units of $\sigma_{\rm vir}$ and $R_{\rm vir}$ respectively. The $\eta$ is defined as $\frac{R_{\rm projected}}{R_{\rm vir}}\frac{|\Delta v|}{\sigma_{\rm vir}}$ and is used to quantify the global environment. $\eta<2$ should mark the limit for a gravitationally bounded system \citep{Shen2017}.}
\end{figure}

\begin{figure*}
	\includegraphics[scale=0.4]{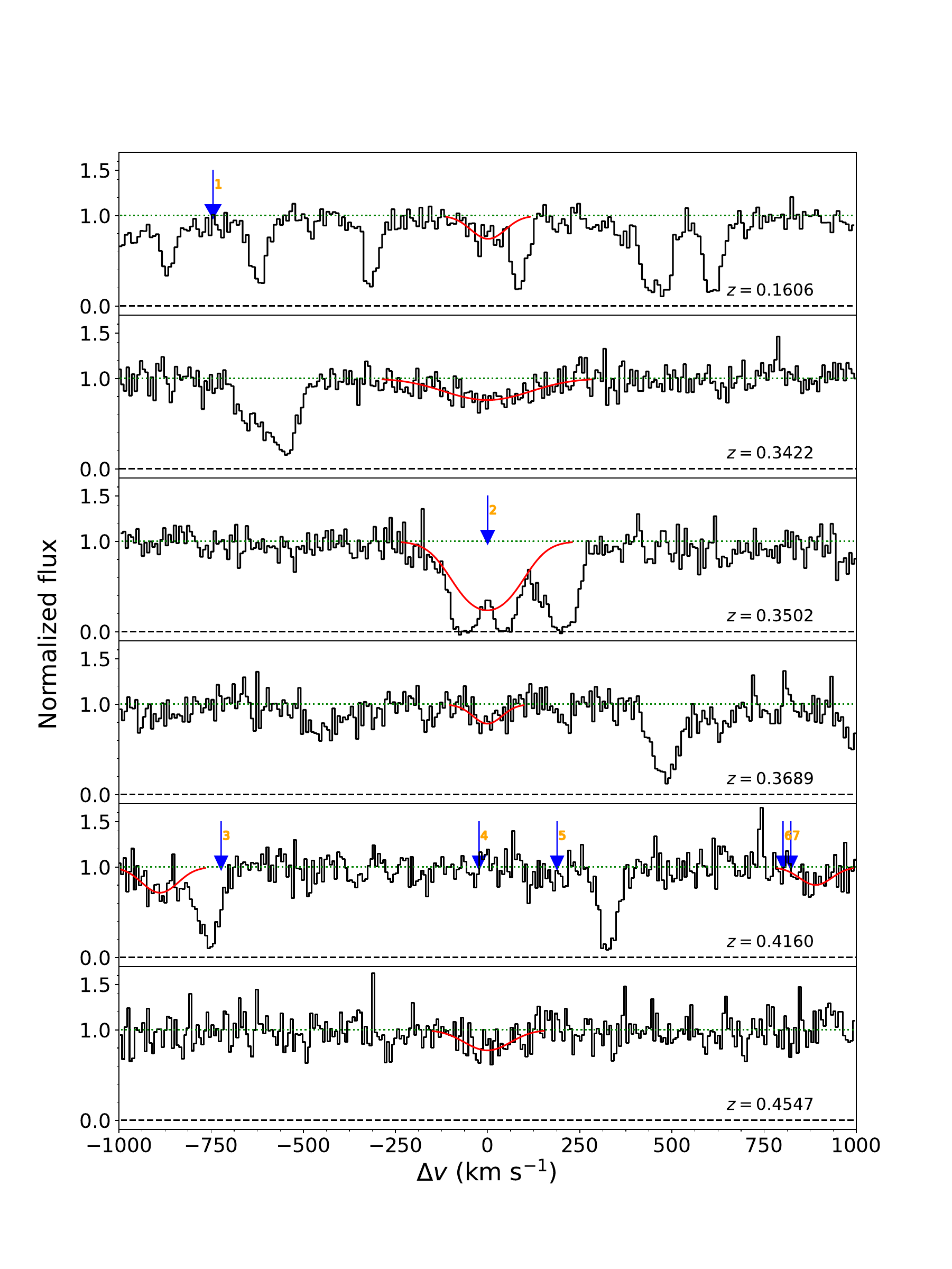}
    \caption{\label{fig:abs}Panels show BLAs in the Q1410 COS spectrum within $\Delta v = \pm 1000$ km s$^{-1}$ from the $z$ of the inter-cluster axes. The panels are ordered by $z$. Arrows with numbers represent the redshifts of nearby galaxies, which are numbered according to Table~\ref{tab:gals} (Col. 1).  
    }
\end{figure*}

In order to discriminate whether a BLA may be associated with one (or more) of these galaxies we consider both their impact parameters and their rest-frame absolute velocity difference $|\Delta v|$. We define $\eta = \frac{R_{\rm projected}}{R_{\rm vir}}\frac{|\Delta v|}{\sigma_{\rm vir}}$ (sometimes called caustic lines) to quantify the global environment.
In Fig.~\ref{fig:blas}, we show the impact parameter of each galaxy versus $|\Delta v|$ with respect to each BLA found at that redshift, in units of $R_{\rm vir}$ and $\sigma_{\rm vir}$ of the corresponding galaxy. Each galaxy-BLA pair is represented by a square coloured by the redshift of the galaxy given by the colour bar. In principle, if a BLA is close to a galaxy both in projection and velocity at values comparable to the  $R_{\rm vir}$ and $\sigma_{\rm vir}$, respectively, it would indicate that the BLA may be have been produced by the galaxy halo rather than by the WHIM. $\eta<2$ should mark the limit for a gravitationally bounded system \citep[][and references therein]{Shen2017}\footnote{This limit were determined in the context of galaxy cluster physics, but we assume the same limit since the involved mechanisms are the same}. For completeness, in Fig.~\ref{fig:abs} we show a portion of the Q1410 spectrum for each one of the inter-cluster axes that present a BLA on its rest-frame. Nearby galaxies are marked with arrows labeled with the IDs given in Table~\ref{tab:gals}.\\



\begin{table}{\large  Galaxies nearby known BLAs}
\centering
\begin{tabular}{lcccccc}
\hline
ID & $z$ & $M_{\rm vir}$ & $R_{\rm vir}$ & I.P. &$\sigma_{\rm vir}$&$\Delta v$\\
 & & $  10^{11}\mathrm{M}_{\odot} $  & kpc&kpc&km s$^{-1}$&km s$^{-1} $\\
(1)  &(2)&(3)&  (4)   &  (5)                      & (6)& (7)\\
\hline
1&0.1577&8.0$\pm$3.9&186$\pm$30&67&136$\pm$35&745\\
2&0.3502&3.0$\pm$1.0&125$\pm$14&199&102$\pm$18&-7\\
3&0.4126&0.8$\pm$0.2&77$\pm$7&91&65$\pm$10&-174\\
4&0.4159&3.4$\pm$1.2&126$\pm$14&199&107$\pm$20&-863\\
5&0.4169&7.2$\pm$3.3&162$\pm$24&173&138$\pm$33&700\\
6&0.4198&3.1$\pm$1.1&122$\pm$13&115&104$\pm$19&80\\
7&0.4199&2.6$\pm$0.9&115$\pm$12&107&98$\pm$17&78\\
\hline
\end{tabular}
\caption{\label{tab:gals}Galactic parameters of galaxies found to be close to the inter-cluster filaments.  The $\Delta v$ of each Galaxy respect to the nearby BLAs is shown in Fig.~\ref{fig:abs}, where each galaxy is labeled according to their ID in this Table. Column (5) corresponds to the physical impact parameter of each galaxy to the Q1410 sight-line in kpc. {Column (7) marks the $\Delta v$ of the galaxy respect to the closest BLA in Fig.~\ref{fig:abs}}. }
\end{table}

\noindent In the following, we give a brief description of each BLA system:
\begin{itemize}
\item The {2} BLAs at $z \approx 0.4160$ seems to be located in a galaxy group environment ({with at least $5$ galaxies)}. There are $2$ galaxies at an impact parameter $\rho/R_{\rm vir} \sim 0.95$ and $|\Delta v|/\sigma_{\rm vir} \sim 0.7${, both} below the $\eta=1$ curve. Thus, we can not rule out that these BLAs may have been produced in the galaxy group halo (as opposed to WHIM). 

\item The BLA at $z \approx 0.3502$ is very close to a strong narrow \hi~system that shows metal enriched gas \citep{Tejos2016}. It also has $|\Delta v|/\sigma_{\rm vir} < 0.5$ to the closest galaxy observed and it is located below the $\eta=1$ curve. This places the absorption close to a galactic halo and its origin is uncertain. {Moreover, our revised analysis of the absorption features in Q1410 (see Appendix~\ref{sec:redo_appendix}) deems this BLA as highly uncertain}. 

\item The BLA at at $z \approx 0.1606$ has $\rho/R_{\rm vir} \sim 0.4$ but $|\Delta v|/ \sigma_{\rm vir} \sim  5.4$. This places this BLA near the limit of $\eta = 2$. However, this latter BLA also shows a nearby galaxy from SDSS at an impact parameter of $\approx 260$\,kpc to the sight-line of the QSO with  $|\Delta v| \approx 800$\,km s$^{-1}$. Moreover, SDSS shows two more galaxies at higher impact parameters ($\sim 450$\,kpc), both at $|\Delta v| \sim 560$\,km s$^{-1}$ (See Section~\ref{sec:sdss_gal}). 

\item The BLAs at $z\approx 0.3422$, $z\approx 0.3689$, and $z\approx 0.4547$  do not show any potential host galaxy to a {luminosity} limit of $L\approx4 \times10^{8}$\,L$_{\odot}$, {corresponding to ${\approx}0.01\,\mathrm{L}_{*}$ \citep{Zucca2006, GAMA2014} (or ${\sim}3\times10^{8} $L$_{\odot}$ for $z\approx0.3422$ and ${\sim}5\times10^{8} $L$_{\odot}$ for $z\approx0.4547$). We note that at those redshifts, our FoV cover impact parameters of $\approx$ 270, 280, and 320\,kpc\footnote{These numbers may vary since our FoV is not symmetrically distributed around Q1410}, respectively, comparable to the virial radius of an L$_{*}$ galaxy. Still, as explained in Section~\ref{sec:sdss_gal}, we have searched for photometric galaxies outside the MUSE FoV from the SDSS data, and found that the presence of such luminous galaxies right outside the FoV is unlikely.} 



\end{itemize}
As mentioned above, a detection limit of $r_{\rm AB} = 25$ mag implies a luminosity {limit of $\sim2-5\times10^{8}$L$_{\odot}$ at $z\approx0.1-0.45$.} We detected 7 galaxies brighter than this luminosity limit at the redshifts of the inter-cluster axes. We used a Schechter luminosity function to estimate the number of fainter galaxies that we could be missing given that we detected 7 galaxies brighter than the luminosity limit. {We used $\alpha=-1.25$ and L$_{*}=3.15 \times 10^{10}$L$_{\odot}$ \citep{GAMA2014} as the parameters for the luminosity function and L$_{\rm min} = 1\times10^{6}$L$_{\odot}$ \citep{Sawala2016} as a conservative luminosity lower limit to integrate the luminosity function (although the existence of even fainter galaxies is possible such as fossils galaxies).} {From this, we estimate that $<1$ faint galaxies may be missing.} Thus, it is unlikely our analysis is significantly affected by an undetected population of faint galaxies. 

{Regarding large-scale structure, we remind the reader that we found evidence of a galaxy group in only $1$ out of the seven inter-cluster axes probed by our MUSE survey (at $z\approx0.461$). The lack of galaxy overdensities in the other $6$, does not rule out the existence of a filamentary structure in them. According to the halo mass function presented in \citet{Reed2007}, the average density of dark matter haloes more massive than $10^{12}$\,$h^{-1}$M$_{\odot}$ is $0.004$\,$(h^{-1}{\rm Mpc})^3$ . The {volume} sampled by MUSE inside the a single filament is about $400$\,kpc $\times 400$\,kpc $\times 6$\,Mpc. Considering an overdensity of factor $\approx 3$ in the filaments with respect to the mean density of the Universe, and the fact that the sight-line passes through $7$ independent inter-cluster axes, we would expect to detect $\approx 0.02$ dark matter haloes more massive than $10^{12} M_{\odot}$ associated to the filaments in the MUSE FoV. This is consistent with what the single structure found in our data, within the Poissonian error.}




\subsection{Revision to the BLAs reported by Tejos et al. 2016}
\label{sec:redo}

Given the intrinsic difficulty for finding and characterizing broad and shallow absorption features in QSO spectra, it is expected that some of the reported BLAs may be subject to 
large systematic uncertainties. In Appendix~\ref{sec:redo_appendix} we have performed independent analyses for quantifying potential systematic effects. We concluded that, with the exception of the putative BLA at $z=0.3502$ (that could be even a narrow \hi), all the other BLAs in inter-cluster axes reported by BLA have systematic uncertainties well below or consistent with the reported statistical uncertainties. Thus, given that the absorption feature at $z=0.3502$ has already been discarded from our `clean' sample of BLAs on the basis of the existence of a potential galaxy counterpart (see Section~\ref{sec:asso}), we expect that the rest of the BLA sample is not much affected by systematic effect in the analysis of \citet{Tejos2016}. In the following, we will use their reported fit parameters for these BLAs as these agree well with our new analyses within statistical uncertainties. 
We note that according to \citet{Tejos2016}, this sample of BLAs is complete down to a rest-frame equivalent width of $W_{\rm r} = 0.039$\,\AA\ (see their figure 5). There are two BLAs in our sample that are close to this limit: the one at $z=0.3689$ (with $W_{\rm r}=0.089\pm0.023$\,\AA) and the one at $z=0.4202$ (with $W_{\rm r}=0.090\pm0.024$\,\AA). The latter is already discarded from our `clean' sample because of the presence of a group of galaxies at a similar redshift, and we have opted to keep the former in.

{In summary,} our analyses indicate {that} $3$ BLAs in inter-cluster axes do not show nearby potential host galaxy halos to stringent {luminosity} limits {and their fit parameters are robust to tests on potential systematic effects (including data reduction, continuum estimation and Voigt profile fitting software)}. We consider these $3$ BLAs to be good WHIM candidates. In the following, we will use their properties to assess the baryon content implied by these BLAs assuming these are genuine WHIM signatures.

\subsection{Density characterization of WHIM}
\label{sec:density}

For the three BLAs that do not show any nearby galaxy we can obtain the neutral hydrogen column density $N_{\rm HI}$ and the Doppler parameter $b$ directly from the observed spectrum \citep{Tejos2016}.
We split the observed Doppler parameter into two different components, thermal ($b_{\rm th}$) and non-thermal ($b_{\rm non-th}$),

\begin{equation}
b_{\rm obs} = \sqrt[]{b_{\rm th}^{2} + b_{\rm non-th}^{2}}
\end{equation}

\noindent The thermal broadening only depends on the temperature ($T$) of the gas,

\begin{equation}
b_{\rm th}=\sqrt[]{\frac{2k_{b}T}{m}} \approx0.129\sqrt[]{\frac{T}{A}}\mathrm{km \, s}^{-1}
\label{eq:bth}
\end{equation}

\noindent where $k_{b}$ is the Boltzmann constant, $m$ is the gas particle mass and $A$ is the atomic weight of the element. For \hi\ equation~\ref{eq:bth} follows \citep[e.g.][]{Richter2006b}:

\begin{equation}
T \approx 60 \left(\frac{b_{\rm th}}{\mathrm{km}\,  \mathrm{s}^{-1}}\right)^{2}\, \mathrm{K}
\label{eq:temp}
\end{equation}


On the other hand, non-thermal broadening mechanisms include turbulence, Hubble flow, line blending, etc. In overdensities like inter-cluster filaments we may expect that turbulence dominates, and we parametrize it as being proportional to the thermal broadening, such that $b_{non-th} \approx b_{turb} \approx \alpha b_{th}$. {This would imply:} 
\begin{equation}
b^2_{\rm obs} \approx b_{\rm th}^2 + b_{\rm turb}^2 \approx b_{\rm th}^2 (1 + \alpha^2)
\label{eq:btot}
\end{equation}

{\noindent Cosmological hydrodynamical simulations from the OWLS project suggest $0 \le \alpha \le 1.3$ with a dependence on \hi\ column density \citep[][see their figure 5]{TepperGarcia2012}. If we restrict to absorption lines with column densities in the range of $10^{13}< N < 10^{14.5}$\,cm$^{-2}$ these simulations suggest $\alpha \approx 0-0.8$. We also note that \citet{Richter2006a} found values of $\alpha \approx 0.5$ from an independent simulation. Observationally, one could estimate $\alpha$ by comparing the observed Doppler parameters in systems with both \hi\ and \ovi, assuming they both come from the same gas in thermal equilibrium \citep{Rauch1997}. \citet{Savage2014} and \citet{Stocke2014} presented a sample with aligned\footnote{$\Delta v < 10$ km s$^{-1}$ between the \hi\ and \ovi\ components} \hi\ and \ovi\ absorbers. Using the different line widths of both components, and given that the thermal Doppler broadening depends on the temperature and on the atomic mass (see Equation~\ref{eq:bth}), they estimated the $b_{\rm non-th}$ contribution for each system. Considering only those systems where $b(\mathrm{HI})_{\rm obs}>40$ km s$^{-1}$ (18 out of their total sample) we studied the distribution for the $\alpha$ parameter in their sample. Their $\alpha$ values are distributed between $0.2<\alpha<2.1$, with an average of $\approx 0.7$ where most of the values ($15/18$) have $\alpha<1$.} 

{Given that the BLAs in our sample do not show \ovi, we can not empirically determinate the non-thermal broadening for our BLAs individually in this manner.}
{Based on these theoretical and empirical studies, in the following we assume a fiducial value of $\alpha = 0.7$, but will leave its dependency explicit in the calculations. From equation~\ref{eq:btot} we then obtain the value of $b_{\rm th}$ given $b_{\rm obs}$. Then we can estimate the temperature of the gas using equation~\ref{eq:temp}.}\\
In order to calculate the total gas column density, it is necessary to know the ionization fraction of the gas,

\begin{equation}
f_{\rm ion}\equiv\frac{N_{\rm HI}+N_{\rm HII}}{N_{\rm HI}}\approx \frac{N_{\rm HII}}{N_{\rm HI}}
\label{eq:fion}
\end{equation}
\noindent i.e., the number of ionized hydrogen per neutral ones. If we take into account a pure collisional ionization equilibrium (CIE) scenario, $f_{\rm ion}$ depends only on the temperature of the gas and can be approximated by the polynomial \citep{SutherlandDopita1993}:

\begin{equation}
\mathrm{log}(f_{\rm ion}) \approx -13.9+5.5\mathrm{log}\left(\frac{T}{\rm K}\right)-0.33\mathrm{log}\left(\frac{T}{\rm K}\right)^{2}
\label{eq:fion_poly}
\end{equation}

\noindent However, \citet{Richter2006a} found that at the typical WHIM densities, photoionization from the UV background also contributes; neglecting it may underestimate the baryon density up to $50\%$. Their combined photoionization plus collisional ionization model suggests a linear relation between $\log(f_{\rm ion})$ and $\log(T)$ as

\begin{equation}
\mathrm{log}(f_{\rm ion}) \approx -0.75+1.25\mathrm{log}\left(\frac{T}{\rm K}\right)
\label{eq:fion_richter}
\end{equation}

for which $f_{\rm ion}$ as a function of $b_{\rm obs}$ and $\alpha$ can be written as:

\begin{equation}
f_{\rm ion} \approx 0.1778 \left(\frac{T}{K}\right)^{1.25} \approx 29.7\left(\frac{b_{\rm obs}/\mathrm{km\, s}^{-1}}{\sqrt[]{1+\alpha^{2}}}\right)^{\frac{5}{2}}  
\label{eq:fion_final}
\end{equation}

Here we use equation~\ref{eq:fion_final} to infer $f_{\rm ion}$ and then calculate the total gas column density $N_{\rm H} \approx N_{\rm HII}$ for each individual BLA simply as $N_{\rm H} \approx f_{\rm ion} N_{\rm HI}$.\\

\noindent We assume a radial volumetric density profile for the inter-cluster filaments parametrized by an exponent $\Gamma$ of the form

\begin{equation}
n_{\rm H}(r) = n_{0}\frac{1}{1+\left(\frac{r}{r_{1/2}}\right)^{\Gamma}}
\label{eq:nHr}
\end{equation}

\noindent where $r$ is the radial distance to the filament axis (i.e. in cylindrical coordinates), $n_{0}$ is the peak density at filament center (except when $\Gamma{=}0$), and $r_{1/2}$ is a characteristic radius such that $n_{H}(r_{1/2}) = n_{0}/2$. Given this radial density profile we can solve

\begin{equation}
N_{\rm H} = \int n_{\rm H} dl
\label{eq:NH}
\end{equation}

\noindent to estimate a mean particle density $\overline{n_{\rm H}}$ inside the filaments given our inferred $N_{\rm H}$. In the following, we use simple models based on two different values of $\Gamma$.

\subsubsection{Uniform density model ($\Gamma = 0$)}
\label{sec:UD}
The simplest case is a model in which the filament is a cylindrical structure with uniform density profile, i.e. $\Gamma=0$ in \ref{eq:nHr}. In this case, 

\begin{equation}
N_{\rm H}=\int_{0}^{L} \frac{n_{0}}{2}dl \equiv \int_{0}^{L}\overline{n_{\rm H}}dl = \overline{n_{\rm H}}L
\label{eq:NH_uniform}
\end{equation}

\begin{figure}
	\includegraphics[width=\columnwidth]{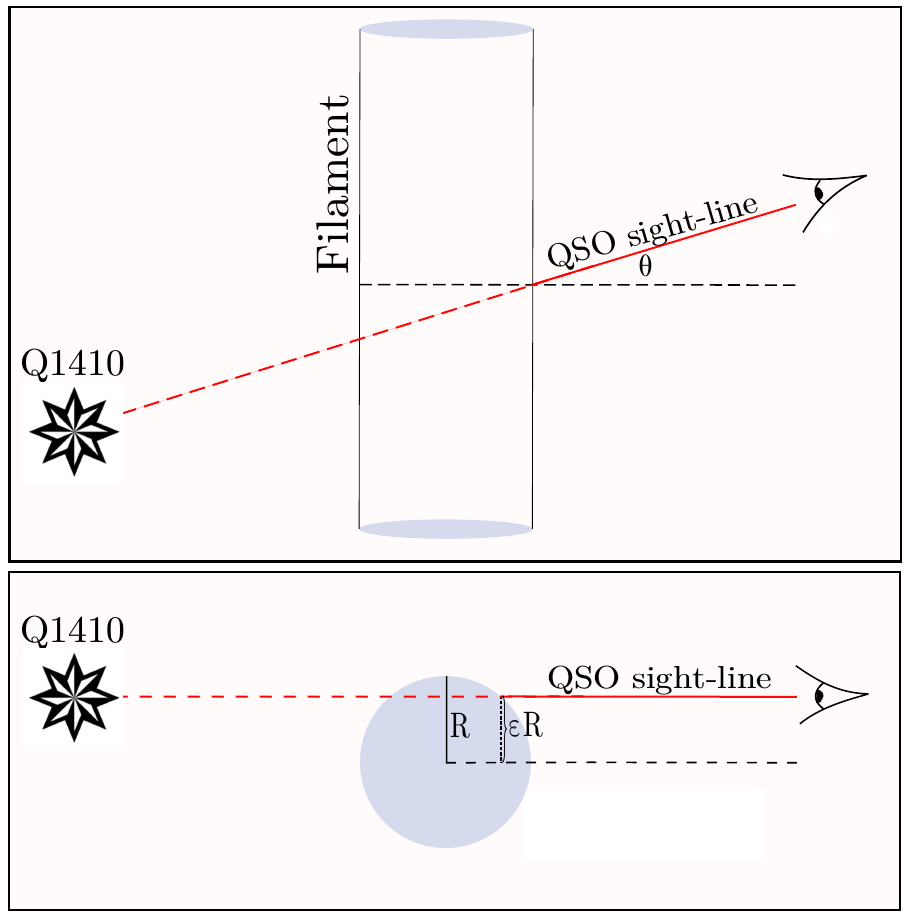}
    \caption{\label{fig:geom}Schematic representation along the sight-line (in red) of the geometry assumed for deriving WHIM density. A filamentary structure is represented by a cylinder. $\theta$ is the angle between the sight-line and the perpendicular axis to the filament. The impact parameter of the sight-line respect to the center is parametrized by the fraction $\epsilon$. }
\end{figure}

\noindent Thus $\overline{n_{\rm H}}$ = $N_{\rm H}/L$ where $L$ corresponds to the distance along the filament intersected by the QSO sight-line. As a first estimation, we can use the average diameter of the cosmic-web filaments of $\approx 6$\,Mpc as found by simulations \citep[e.g][]{GonzalezPadilla2010, AragonCalvo2010,Cautun2014}. The true relation between $L$ and the radius of the filament is uncertain, and it depends on the angle of incidence ($\theta$) and the impact parameter relative to the radius of the filament ($\equiv \epsilon R$) of the sight-line to the center (e.g. see Figure ~\ref{fig:geom} for an schematic). If $\theta \neq 0$ and $\epsilon=0$, i.e. the sight-line is not perpendicular to the filament but it passes through its center, $L = 2R/\cos(\theta)$. On the other hand, if $\theta = 0$ and $\epsilon \neq 0$, i.e. the sight-line is perpendicular to the filament but has an impact parameter of $\epsilon R$, $L = 2 \sqrt[]{1-\epsilon^{2}}R$ (see Fig.~\ref{fig:geom}). Thus, the mean density value for each BLA can be estimated as

\begin{equation}
\overline{n_{\rm H}}=\frac{N_{\rm H}}{L} = \frac{f_{\rm ion} N_{\rm HI}\mathrm{cos}(\theta)}{2\sqrt[]{1-\epsilon^{2}}R}
\label{eq:nh_uniform}
\end{equation}

For simplicity, we assume a fiducial case where the sight-lines are all perpendicular and intersect the filaments at their centers. By averaging the $\overline{n_{\rm H}}$ for the three BLAs with no galaxies nearby we obtain 
\begin{equation}
\langle \overline{n_{\rm H}} \rangle =5.9{\times}10^{-6} \left( \frac{1-\epsilon^{2}}{1} \right) ^{-\frac{1}{2}} \left( \frac{\mathrm{cos}(\theta)}{1} \right) \left(\frac{ 1+\alpha^{2} }{1.49} \right)^{-\frac{5}{4}} \mathrm{cm}^{-3}
\label{eq:ave_nh_uniform}
\end{equation}

\noindent where we have left explicit the dependencies on our parameterizations. 

\subsubsection{Radial density profile model $\propto r^{-2}$ ($\Gamma = 2$)}
\label{sec:RD}

As a more realistic approach, here we try a density profile for the inter-cluster filaments with a core and that follows $n_{\rm H}(r) \propto r^{-2}$ at large distances as suggested by simulations \citep[e.g.][]{AragonCalvo2010}

\begin{equation}
n_{\rm H} = n_{\rm 0} \frac{1}{1+\left(\frac{r}{r_{1/2}}\right)^{2}}
\end{equation}

\noindent Assuming the same geometry as in the previous scenario, we integrate the density over a single sight-line to calculate the core density $n_{0}$

\begin{equation}
n_{0} = \frac{N_{\rm H}  \gamma  \cos(\theta)}{2 r_{1/2}^{2} \mathrm{tan}^{-1}(a/\gamma)}
\label{eq:n0}
\end{equation}

\noindent where $\gamma$ and $a$ are geometrical parameters defined as follows
\begin{equation}
\gamma \equiv R \ \sqrt[]{\epsilon^{2}+\left(\frac{r_{1/2}}{R}\right)^{2}}
\end{equation}
and
\begin{equation}
a \equiv R \ \sqrt[]{1-\epsilon^2}
\end{equation}

\noindent and correspond to the typical size and typical core size of a filament. According to the results presented in \citet{Cautun2014}, we use a value of $r_{1/2} = 2$\,Mpc and integrate the density profile until it becomes approximately flat, i.e. at $R \approx 6$\,Mpc. Using the three good WHIM candidate BLAs in our sample, we can calculate a typical core density $\overline{n_{0}}$ and average the density profiles over the transversal area of the filament to obtain the averaged mean particle density:
\begin{equation}
\langle \overline{n_{\rm H}} \rangle = \overline{n_{0}}\int_A \frac{1}{1+(\frac{r}{r_{1/2}})^{2} }dA
\end{equation}

\noindent From this, and using our fiducial values of $a=6$ Mpc and $\gamma=2$ Mpc we obtained a value of:

\begin{equation}
\overline{n_{0}}=7.1{\times} 10^{-6} \left( \frac{\mathrm{tan}^{-1}(a/\gamma)}{1.25}\right)^{-1} \left( \frac{\gamma}{2~\mathrm{Mpc}}\right) \left( \frac{\mathrm{cos}(\theta)}{1} \right)\left(\frac{ 1+\alpha^{2} }{1.49} \right)^{-\frac{5}{4}} \mathrm{cm}^{-3}
\end{equation}

\noindent for an average total integrated density of

\begin{equation}
\langle \overline{n_{H}} \rangle=2.9{\times} 10^{-6} \left( \frac{\mathrm{tan}^{-1}(a/\gamma)}{1.25}\right)^{-1} \left( \frac{\gamma}{2~\mathrm{Mpc}}\right) \left( \frac{\mathrm{cos}(\theta)}{1} \right)\left(\frac{ 1+\alpha^{2} }{1.49} \right)^{-\frac{5}{4}} \mathrm{cm}^{-3}
\label{eq:ave_nh_2}
\end{equation}

\noindent Table~\ref{tab:BLAs} summarizes the main findings adopting our fiducial values for the individual WHIM candidate BLAs. For completeness, in Table~\ref{tab:BLAs} we have also included the corresponding calculations for those BLAs that show nearby galaxies (marked with `y' in the last column). Note that for the latter, the presence of nearby galaxies complicates the interpretation on its origin but does not necessarily rule out the possibility that these are genuine WHIM absorption. In fact we do not see an important quantitative difference in the inferred densities for good and uncertain WHIM candidates. {We obtained a mean hydrogen particle density of $5.9 \pm 4.1 \times 10^{-6}$\,cm$^{-3}$ in the uniform density scenario (Sect.~\ref{sec:UD}) using the samples of BLAs that do not show nearby galaxies. If we use only the BLAs that could be potentially associated with a galaxy, we obtain a mean particle density of $4.9\pm4.2 \times 10^{-6}$\,cm$^{-3}$. Similarly, in the radial density profile scenario we obtained a mean particle density of $2.9 \pm 2.0 \times 10^{-6}$\,cm$^{-3}$ (Sect.~\ref{sec:RD}). If we consider now the BLAs with uncertain origin only, we obtain a value of $2.4\pm2.1 \times 10^{-6}$\,cm$^{-3}$.}



\begin{table*}{\large Parameters derived from BLAs}
\centering
\begin{tabular}{lccccccccc}
\hline
ID&Redshift&$b$&\multicolumn{2}{c}{Column densities}&$\mathrm{log} f_{\rm ion}$&Temperature&\multicolumn{2}{c}{Gas density}&Has galaxy\\
&&&&&&&$\Gamma=0$&$\Gamma=2$&nearby?\\
          & & km s$^{-1}$ & $\mathrm{log}(N_{\rm HI}/\mathrm{cm}^{-2})$  & $\mathrm{log}(N_{\rm HII}/\mathrm{cm}^{-2})$&  &log($T$/K)  & \multicolumn{2}{c}{$\mathrm{log}(\overline{n_{\rm H} }/\mathrm{cm}^{-3})$}&     \\
(1)  &(2)&(3)&  (4)   &  (5)                      & (6)   &  (7)  &(8)&(9)&(10)\\
\hline
1&0.1606&59$\pm$22&13.40$\pm$0.11&19.1$\pm$0.4&5.68$\pm$0.40&5.1$\pm$0.3&-6.2$\pm$0.4&-6.5$\pm$0.4&y\\
2&0.3422&153$\pm$19&13.75$\pm$0.05&20.5$\pm$0.1&6.72$\pm$0.13&6.0$\pm$0.1&-4.8$\pm$0.1&-5.1$\pm$0.1&n\\
3&0.3502&97$\pm$10&14.29$\pm$0.09&20.5$\pm$0.1&6.22$\pm$0.11&5.6$\pm$0.1&-4.8$\pm$0.1&-5.1$\pm$0.1&y\\
4&0.3689&50$\pm$18&13.25$\pm$0.11&18.8$\pm$0.4&5.50$\pm$0.39&5.0$\pm$0.3&-6.5$\pm$0.4&-6.8$\pm$0.4&n\\
5&0.4118&62$\pm$18&13.47$\pm$0.09&19.2$\pm$0.3&5.74$\pm$0.32&5.2$\pm$0.3&-6.1$\pm$0.3&-6.4$\pm$0.3&y\\
6&0.4202&56$\pm$20&13.25$\pm$0.12&18.9$\pm$0.4&5.63$\pm$0.39&5.1$\pm$0.3&-6.4$\pm$0.4&-6.7$\pm$0.4&y\\
7&0.4547&81$\pm$18&13.46$\pm$0.08&19.5$\pm$0.3&6.03$\pm$0.24&5.4$\pm$0.2&-5.8$\pm$0.3&-6.1$\pm$0.3&n\\
\hline
\end{tabular}
\caption{\label{tab:BLAs}Characterization of each BLA. The errors associated to the redshifts in the column (2) are ${\sim}{\pm}10$ km s$^{-1}$. The mean density $\overline{n_{\rm H}}$ was calculated for both scenarios proposed in Sect.~\ref{sec:density} for all BLAs.\ Column (10) marks as `n' the absorptions that are likely produced by WHIM according to Sect.~\ref{sec:asso}}
\end{table*}

\section{Discussion}
\label{sec:discussion}

\subsection{The relation between BLAs and WHIM}

Based on our blind MUSE survey, we have a first estimation of the fraction of BLAs in inter-cluster filaments that may represent genuine WHIM signatures based on the lack of nearby galaxies to stringent luminosity limits. {We find this fraction to be ${\sim}40\%$ (3/7)}. Despite an analysis of the luminosity function, we cannot rule out the presence of fainter or dust-enshrouded galaxies. 
How this number relates to BLAs discovered in blind absorption surveys is particularly relevant for observational studies focused on the WHIM. Although properly addressing this question is beyond the scope of this paper, we note that one could repeat the experiment presented here in non-targeted QSO sightlines and undertake an empirical comparison. 



\subsection{The baryon density in inter-cluster filaments}
\label{sec:OmegaInFil}
Here we provide an estimation of the implied baryon density in inter-cluster filaments, $\Omega_{\rm b}^{\rm fil}$. The results presented in Section ~\ref{sec:density} correspond to the typical volumetric densities implied by our 3 BLAs that are good WHIM candidates.  In order to estimate the corresponding baryon density, we need to estimate the relative volume occupied by these large-scale filaments in the Universe $V_{\rm fil}$. Thus,:

\begin{equation}
\Omega_{\rm b}^{\rm fil} = \frac{8\pi G m_{\rm H}}{3H(z)^{2}(1-Y)} \langle \overline{n_{\rm H}} \rangle V_{\rm fil}
\label{eq:Obar_fil}
\end{equation}

\noindent where $G$ is the gravitational constant, $m_{\rm H}$ is the hydrogen mass, $H(z)$ is the Hubble parameter, and $Y$ is the baryonic mass fraction in Helium. In the following we use a typical volume $V_{\rm fil} \sim 6\%$ as inferred by cosmological simulations \citep[e.g.][]{Cautun2014}. {We note that this volume fraction may not be consistent with a uniform density model, given that such number comes from simulations with a non-uniform density profile. Nevertheless, we have kept this number fixed as a fiducial value, but our results will be expressed explicitly on $V_{\rm fil}$.}


\subsubsection{Uniform density model}\label{sec:diss:uniform}
In the scenario of filaments with uniform density (see Sect.~\ref{sec:UD}), assuming $\theta{=}0$, $\epsilon{=}0$ and $\alpha =0.7$ as fiducial values, we estimated a mean gas particle density of $\langle \overline{n_{\rm H}} \rangle \approx 5.9 {\times} 10^{-6} \: \mathrm{cm}^{-3}$.
According to equations \ref{eq:ave_nh_uniform} and \ref{eq:Obar_fil} we have

\begin{equation}
\Omega_{\rm b}^{\rm fil} \approx 0.06\left(\frac{\langle \overline{n_{\rm H}} \rangle }{5.9 x 10^{-6}\mathrm{cm}^{-3}}\right)\left(\frac{V_{\rm fil}}{0.06}\right)\left(\frac{1-Y}{0.76} \right) ^{-1}
\end{equation}

\noindent {giving us a value of $\Omega_{\rm b}^{\rm fil} \approx 0.06 \pm 0.04$,} i.e. {somewhat} larger than the total baryon density expected $\Omega_{\rm bar} \approx 0.048$ \citep[e.g.][]{Planck2015}
but consistent within the errors. {The error in this estimations comes from the standard error of the mean density of our sample.}

{With the aim at reducing the statistical uncertainties, here we also include the densities of the BLAs reported in \citet{Wakker2015}. The authors reported a total of $5$ BLAs, $4$ of them in sight-lines within $540$\,kpc to the axis of the filament and $1$ of them at an impact parameter of $\sim3$\,Mpc. We added these $5$ BLAs to our sample, and calculated the corresponding $\langle \overline{n_{\rm H}} \rangle$ in the same manner as in Sect.~\ref{sec:density}.\footnote{We note that the BLAs reported in \citet{Wakker2015} arise from the same filamentary structure instead of independent ones.} Combining their and our sample of BLAs we obtain a somewhat better constrained value of $\Omega_{\rm b}^{\rm fil} \approx 0.04 \pm 0.02$.}

\subsubsection{Radial density model $\propto r^{-2}$}

Alternatively, in the scenario of a density profile $\propto r^{-2}$ (see Section~\ref{sec:RD}) and using the same fiducial values for the main parameters as in the previous section, we estimated a mean gas peak density of $\overline{n_{0}}{=}7.1{\times}10^{-6}$\,cm$^{-3}$. By averaging the density profile over the transversal area of the filament, we obtained an overall mean gas particle density of   
$\langle \overline{n_{\rm H}}\rangle {=} 2.9 {\times} 10^{-6}$\,cm$^{-3}$. Using equations \ref{eq:ave_nh_2} and \ref{eq:Obar_fil} we obtain an alternative value for $\Omega_{\rm b}^{\rm fil}$ as

\begin{equation}
\Omega_{\rm b}^{\rm fil} \approx 0.03\left(\frac{\langle \overline{n_{\rm H}}\rangle}{2.9 x 10^{-6}\mathrm{cm}^{-3}}\right)\left(\frac{V_{\rm fil}}{0.06}\right)\left(\frac{1-Y}{0.76}\right)^{-1}
\end{equation}

\noindent {This new fiducial value $\Omega_{\rm b}^{\rm fil} \approx 0.03 \pm 0.02$ is somewhat lower than $\Omega_{\rm bar}\approx 0.048$, {but subject} to large statistical and systematic uncertainties. {If we include in our sample the BLAs reported in \citet{Wakker2015} (see Section~\ref{sec:diss:uniform}) we obtain a somewhat better constrained value of $\Omega_{\rm b}^{\rm fil} \approx 0.02 \pm 0.01$, consistent with expected value of $0.4 \times \Omega_{\rm bar} \approx 0.02$}.}





\section{Summary and Conclusions}
\label{sec:conc}

In this paper we used VLT/MUSE to perform a blind galaxy survey around a unique QSO whose sightline may be passing through $7$ inter-cluster filaments as presented in \citet{Tejos2016}. In particular, we focus on the presence or lack of galaxies within $\Delta v = \pm 1000$\,km s$^{-1}$ from each of the $7$ broad \hi~Ly$\alpha$ absorption (BLAs) found at these inter-cluster redshifts in order to determine their origin.\\


We detected $77$ sources and characterized the redshift of $52$  of them. We reached $100\%$ characterization completeness down to magnitude $r = 23$\,mag, and $\approx 75\%$ completeness down to magnitude $r = 25$\,mag. We found that $4$ of the BLAs showed nearby galaxies, for which the origin of the BLAs {is uncertain}. These include a galaxy group at $z \approx 0.416$ and a potential galaxy halo BLA at $z\approx 0.35$ {(not well constrained by the data, see Appendix~\ref{sec:redo_appendix})}. On the other hand, we found $3$ BLAs that do not show any galaxy nearby to stringent luminosity limits. The lack of a nearby galaxy implies that they may be produced by the long sought after warm-hot intergalactic medium (WHIM), and would mean that a {significant fraction ($\sim 40\%$) of the BLAs detected between cluster-pairs (where the existence of a filamentary structure can be expected)}, may be {directly} tracing WHIM.\\

Assuming these BLAs are genuinely produced by the WHIM, we estimated the mean gas particle density assuming two different density profile models for the filaments themselves. First, we used an uniform density profile scenario with a set of assumptions on the geometry and the broadening mechanisms involved, and estimated a rough mean gas density of $\langle\overline{n_{H}}\rangle {\sim}(5.9 \pm 4.1) \times 10^{-6}$\,cm$^{-3}$. This value implies an unrealistically large $\Omega_{\rm bar}^{\rm fil}{\approx}0.06 \pm 0.04$ ,i.e. larger than the expected total baryon density of $\Omega_{\rm bar}{\approx}0.048$ \citep{Planck2015} but still consistent within errors. In the second scenario, we assumed a radial density profile of the form $n_{\rm H}(r){\propto} r^{-2}$ \citep[as suggested by simulations; e.g.][]{AragonCalvo2010}, which led us to estimate a mean gas particle density of $\langle \overline{n_{\rm H}} \rangle {\sim} (2.9 \pm 2.0) \times 10^{-6}$\,cm$^{-3}$ and a corresponding $\Omega_{\rm bar}^{\rm fil}{\approx}0.03 \pm 0.02$. This value is similar to the expected $0.4\Omega_{\rm bar} {\approx} 0.02$ that may be in a WHIM state. {Including the BLAs presented in \citet{Wakker2015} into our sample we obtain somewhat better constrained values for $\Omega_{\rm bar}^{\rm fil}$ of $0.04 \pm 0.02$ and $0.02 \pm 0.01$ for the uniform density and radial density model, respectively.}

We emphasize that these estimations are subject to large statistical and systematic uncertainties, owing to our small sample of {BLAs ($3$ and $3+5$)} and to the intrinsic uncertainties of our {assumed} geometrical parameters. Furthermore, the relationship between observed Doppler $b$ parameters and gas-phase temperature is affected by poorly-constrained physical processes (e.g. turbulence). The results presented here support the hypothesis that inter-cluster filaments host a significant amount of baryons, enough to close the baryon budget in the low-$z$ Universe {but larger samples need to be analyzed for conclusive results.}



\section*{Acknowledgements}
Our results are based on observations collected at the European Organisation for Astronomical Research in the Southern Hemisphere under ESO programme 094.A-0575(C). Some of the data presented in this paper were obtained from the NASA/ESA  Hubble Space Telescope under programme GO 12958, obtained at the Space Telescope Science Institute and from the Mikulski Archive for Space Telescopes (MAST). STScI is operated by the Association of Universities for Research in Astronomy, Inc., under NASA contract NAS5-26555. {We thank the referee, John Stocke, for providing valuable comments and criticism that improved the paper.}  I.P. and N.T. acknowledge support from {\it CONICYT PAI/82140055.}\\
We thank contributors to SciPy, Matplotlib, Astropy \citep{astropy} and the PYTHON programming language; the free and open-source community; and the NASA Astrophysics Data System for software and services. We also thank contributors to linetools \citep{linetools} and {\sc PyMUSE} \citep{PyMUSE}, both open-source PYTHON packages recently developed and used in this work.








\appendix
\section{Uncharacterized sources}
\label{sec:noidsources}
\begin{table*}{\large Sources not characterized in our survey}
\centering
\begin{tabular}{lcccccr}
\hline
ID&Object&RA&DEC&Impact Parameter&$r_{\rm AB}$&reliability\\
          & & J2000 & J2000   &    (arcsecs)&        &   \\
(1)  &(2)&(3)&  (4)   &  (5)                      & (6)   &  (7) \\
\hline
1&J141038.12+230441.3&212.65883&23.07814&6.24&-&d\\
2&J141038.57+230440.7&212.66071&23.07797&6.42&24.27&d\\
3&J141038.54+230453.0&212.66058&23.08139&6.91&24.79&d\\
4&J141038.17+230435.9&212.65904&23.07664&10.95&25.36&d\\
5&J141037.60+230443.3&212.65667&23.07869&11.10&-&d\\
6&J141037.51+230445.8&212.65629&23.07939&11.89&24.57&d\\
7&J141039.24+230453.5&212.66350&23.08153&13.90&-&d\\
8&J141039.15+230456.4&212.66312&23.08233&14.62&-&d\\
9&J141038.38+230430.3&212.65992&23.07508&16.20&-&d\\
10&J141039.03+230430.7&212.66262&23.07519&18.24&24.21&d\\
11&J141038.55+230425.6&212.66062&23.07378&21.05&24.69&d\\
12&J141039.98+230454.5&212.66658&23.08181&23.61&-&d\\
13&J141037.24+230505.8&212.65517&23.08494&24.81&25.57&d\\
14&J141039.87+230425.6&212.66612&23.07378&29.42&27.38&d\\
15&J141036.36+230426.5&212.65150&23.07403&34.20&23.96&d\\
16&J141036.33+230507.5&212.65137&23.08542&35.12&-&d\\
17&J141036.65+230419.5&212.65271&23.07208&35.95&-&d\\
18&J141040.73+230508.3&212.66971&23.08564&39.19&24.96&d\\
19&J141038.38+230526.0&212.65992&23.09056&39.50&-&d\\
20&J141036.28+230418.2&212.65117&23.07172&40.41&-&d\\
21&J141041.09+230503.6&212.67121&23.08433&41.25&25.22&d\\
22&J141040.93+230508.3&212.67054&23.08564&41.51&26.30&d\\
23&J141037.52+230526.5&212.65633&23.09069&41.68&25.27&d\\
24&J141041.61+230437.6&212.67337&23.07711&45.59&25.17&d\\
25&J141039.53+230532.5&212.66471&23.09236&48.71&24.06&d\\
\hline
\end{tabular}
\caption{\label{tab:noids}List of the sources that could not be characterized. Sources where $r$ is undefined were not detected by SExtractor and we manually included them in the survey.}
\end{table*}

{Table~\ref{tab:noids} lists photometric sources identified in the MUSE dataset but for which no redshift solution was found.}

\section{K-correction}\label{sec:kcorr}

We have calculated the photometric K-correction in the $r$ filter for galaxies in our survey as follows.
We used empirical templates presented in \citet{Coleman1980} extended to blue wavelengths according to \citet{BruzualCharlot2003} models to calculate the predicted $r-i$ color and observed $r_{\rm AB}$ at different redshifts for different spectral types from E to Im. We obtained a monotonic relation between $r-i$ color and K-correction for each redshift bin. We then empirically calculated the $r-i$ color for the galaxies in our survey with known redshifts and interpolated this relation to obtain their corresponding K-correction. 

\section{SDSS galaxies}
\label{sec:SDSS}
\begin{table}{\large SDSS spectroscopic galaxies}
\centering
\begin{tabular}{lcccr}
\hline
\multicolumn{2}{c}{Object}&$r$&Impact parameter&$z$\\
          RA&DEC & &kpc&\\
\hline
14:10:33.00&+23:05:44.87&17.04&264&0.1579\\
14:10:46.43&+23:02:49.87&17.12&456&0.1584\\
14:10:50.97&+23:05:13.20&16.72&494&0.1580\\

\hline
\end{tabular}
\caption{\label{tab:SDSS}List of the SDSS spectroscopic galaxies outside the MUSE FoV and within $500$\,kpc from the QSO sightline.}
\end{table}
{Table~\ref{tab:SDSS} lists SDSS spectroscopic galaxies from the SDSS outside our MUSE FoV up to impact parameters of $\approx$500\,kpc.}

\section{Redmonster redshifts measurements}
\label{sec:appims}

\begin{figure*}
	\includegraphics[scale=0.16]{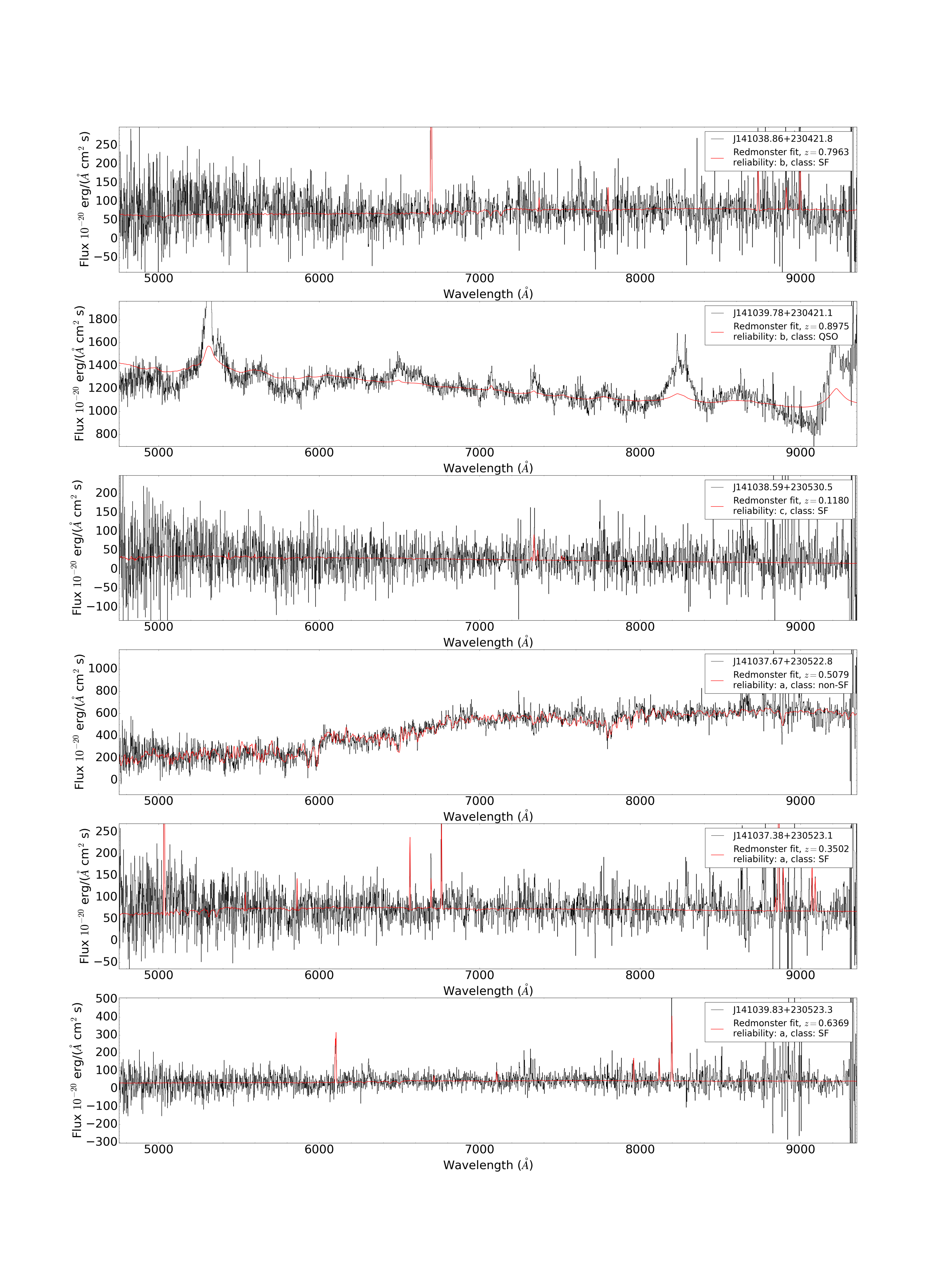}
    \caption{{Spectra characterized using Redmonster software}}
    \label{fig:Redmonster1}
\end{figure*}

\begin{figure*}
	\includegraphics[scale=0.16]{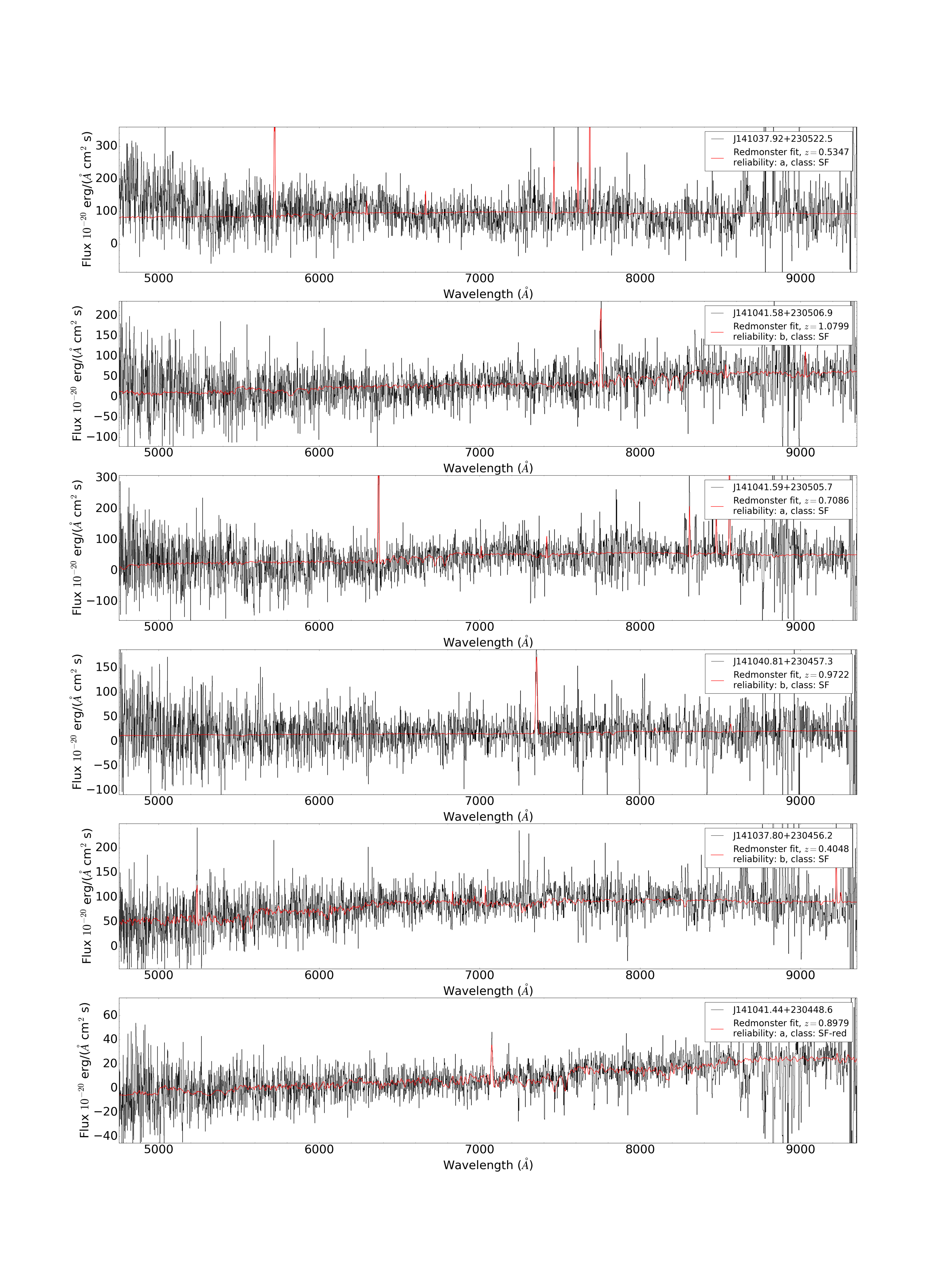}
    \caption{{Spectra characterized using Redmonster software}}
    \label{fig:Redmonster2}
\end{figure*}

\begin{figure*}
	\includegraphics[scale=0.16]{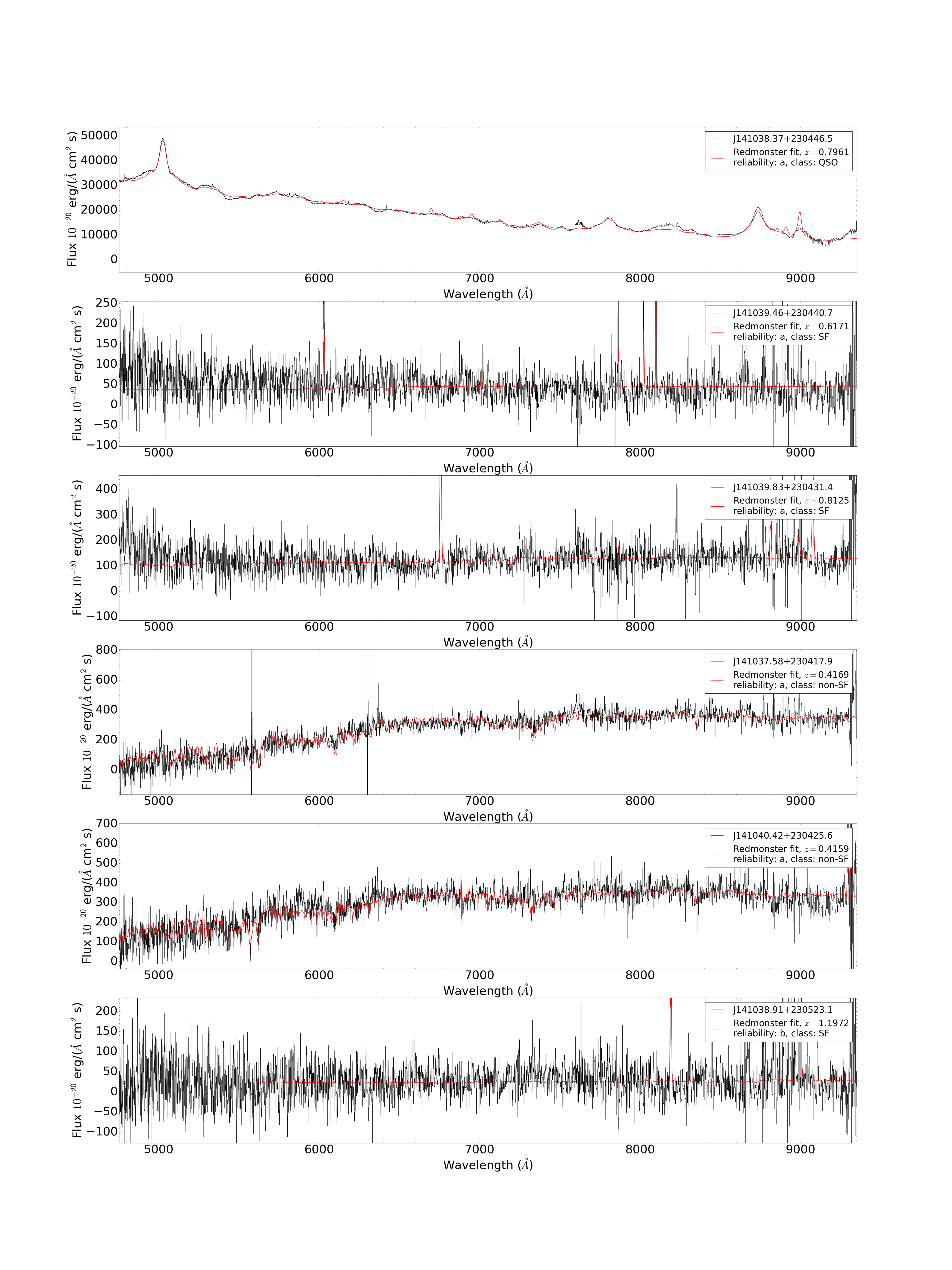}
    \caption{{Spectra characterized using Redmonster software}}
    \label{fig:Redmonster3}
\end{figure*}

\begin{figure*}
	\includegraphics[scale=0.16]{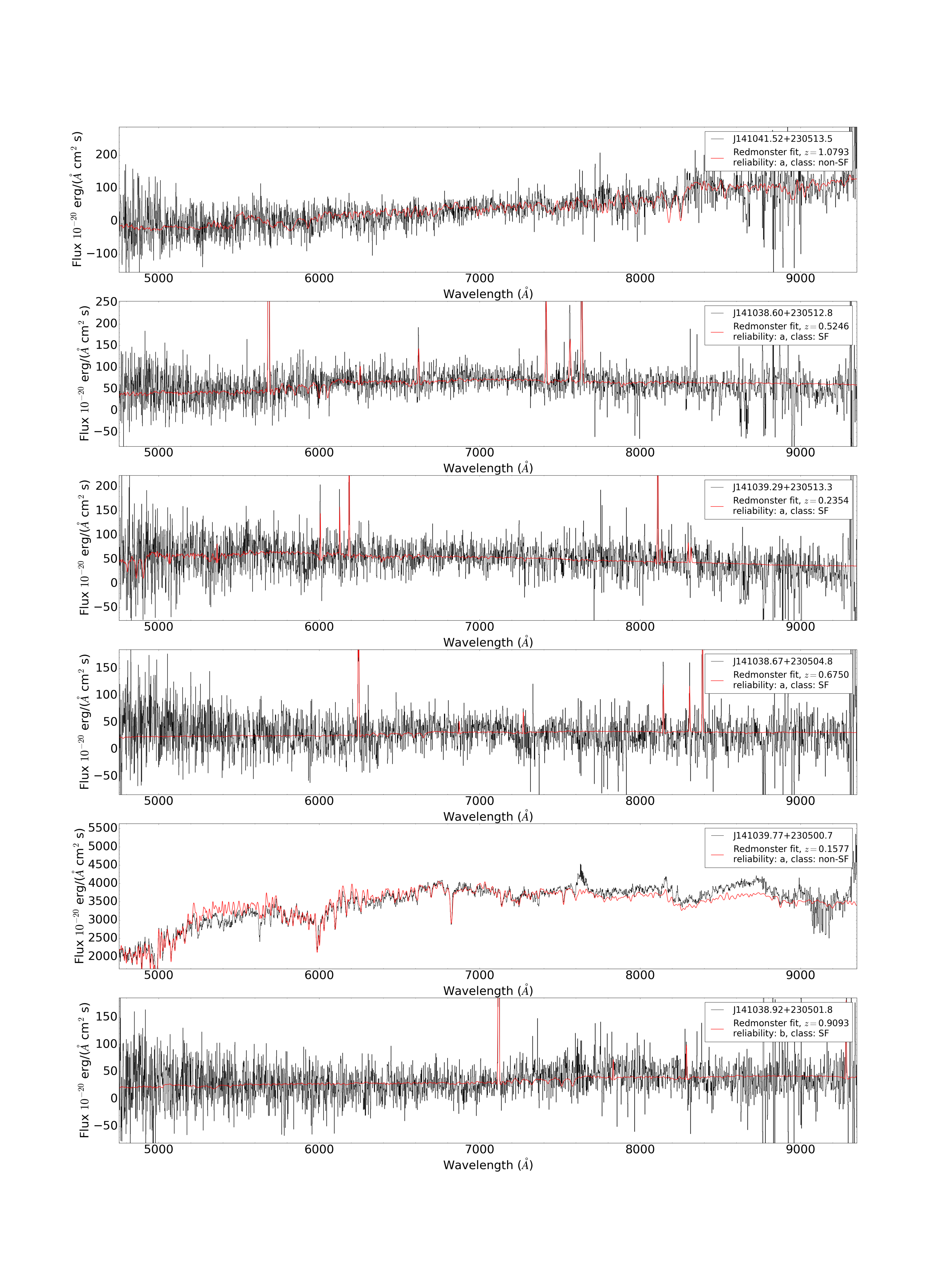}
    \caption{{Spectra characterized using Redmonster software}}
    \label{fig:Redmonster4}
\end{figure*}

\begin{figure*}
	\includegraphics[scale=0.16]{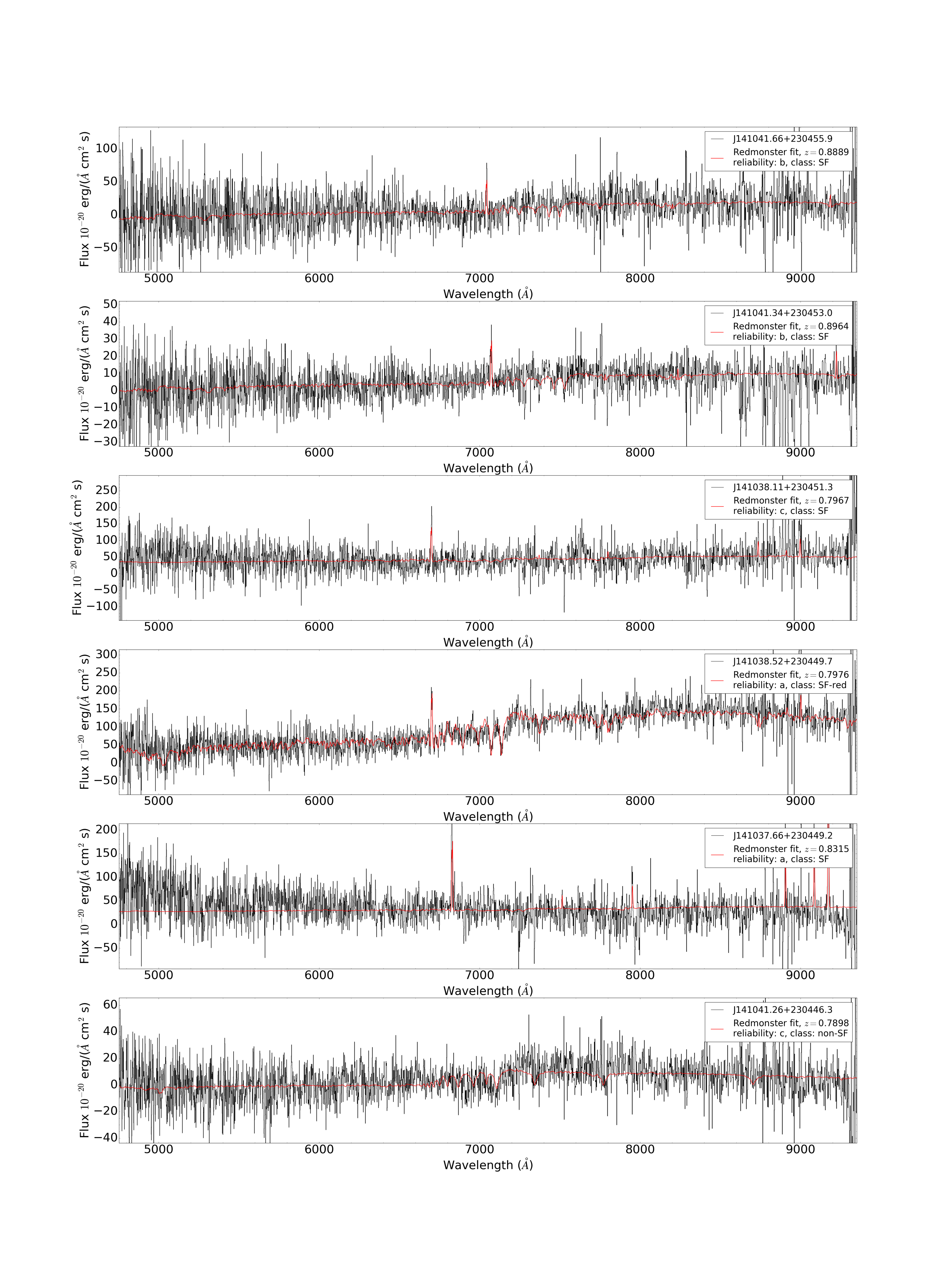}
    \caption{{Spectra characterized using Redmonster software}}
    \label{fig:Redmonster5}
\end{figure*}

\begin{figure*}
	\includegraphics[scale=0.16]{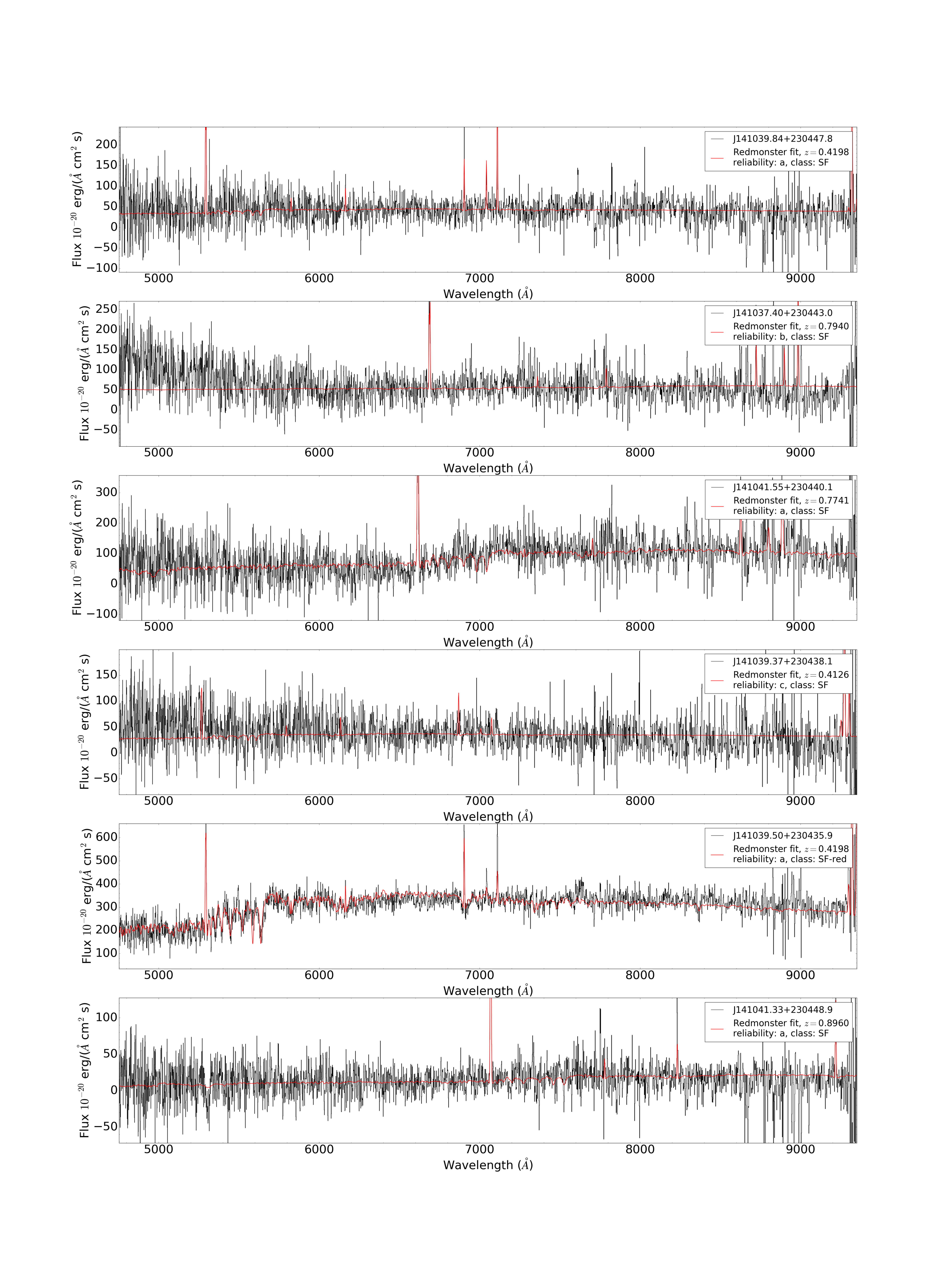}
    \caption{{Spectra characterized using Redmonster software}}
    \label{fig:Redmonster6}
\end{figure*}
\begin{figure*}
	\includegraphics[scale=0.16]{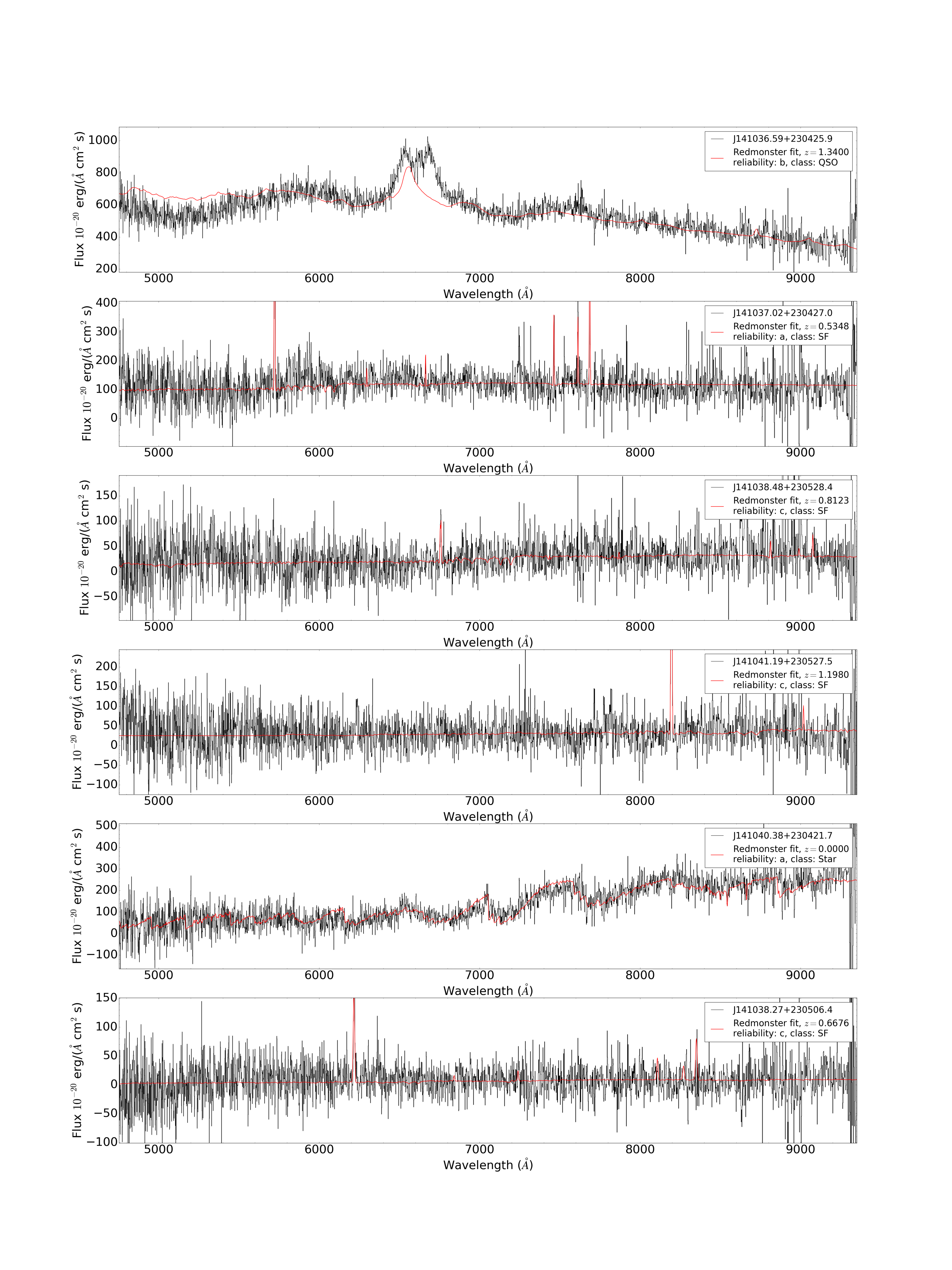}
    \caption{{Spectra characterized using Redmonster software}}
    \label{fig:Redmonster7}
\end{figure*}

{Figs.~\ref{fig:Redmonster1},~\ref{fig:Redmonster2},~\ref{fig:Redmonster3},~\ref{fig:Redmonster4},~\ref{fig:Redmonster5},~\ref{fig:Redmonster6} and ~\ref{fig:Redmonster7} shows the spectra of identified sources in the MUSE data that were characterized using Redmonster software.}

\section{Re-analysis of BLAs reported by Tejos et al. 2016}
\label{sec:redo_appendix}
{In order to test for potential systematic errors in the characterization of BLAs by \citet{Tejos2016}, here we have revisited their identifications and fit parameters. In particular, we repeated their analyses using the following:}

\begin{itemize}
{
\item {\it Different data reduction} For our comparison we used the reduced Q1410 HST/COS data provided by the MAST HST archive\footnote{\url{https://archive.stsci.edu/hst/}} (referred to as `new data') as opposed to the custom reduction done by \citet[][referred to as `old data']{Tejos2016}. We note that the `old data' has slightly higher signal-to-noise than the `new data'.

\item {\it Different continuum level estimation} For our comparison we used an independent estimation of the continuum level obtained from the `new data' (referred to as `new continuum') as opposed to the old continuum estimation described by \citet[][referred to as `old continuum']{Tejos2016}.

\item {\it Different Voigt profile fitting software} For our comparison we used {\sc veeper}\footnote{Mainly developed by J. Burchett; available at
\url{https://github.com/jnburchett/veeper}.} as opposed to {\sc vpfit}\footnote{Developed by R.F.Carswell and J.K.Webb; available at \url{http://www.ast.cam.ac.uk/\~rfc/vpfit.html}.} used by \citet{Tejos2016}.
}
\end{itemize}

\begin{figure*}
	\includegraphics[scale=.6]{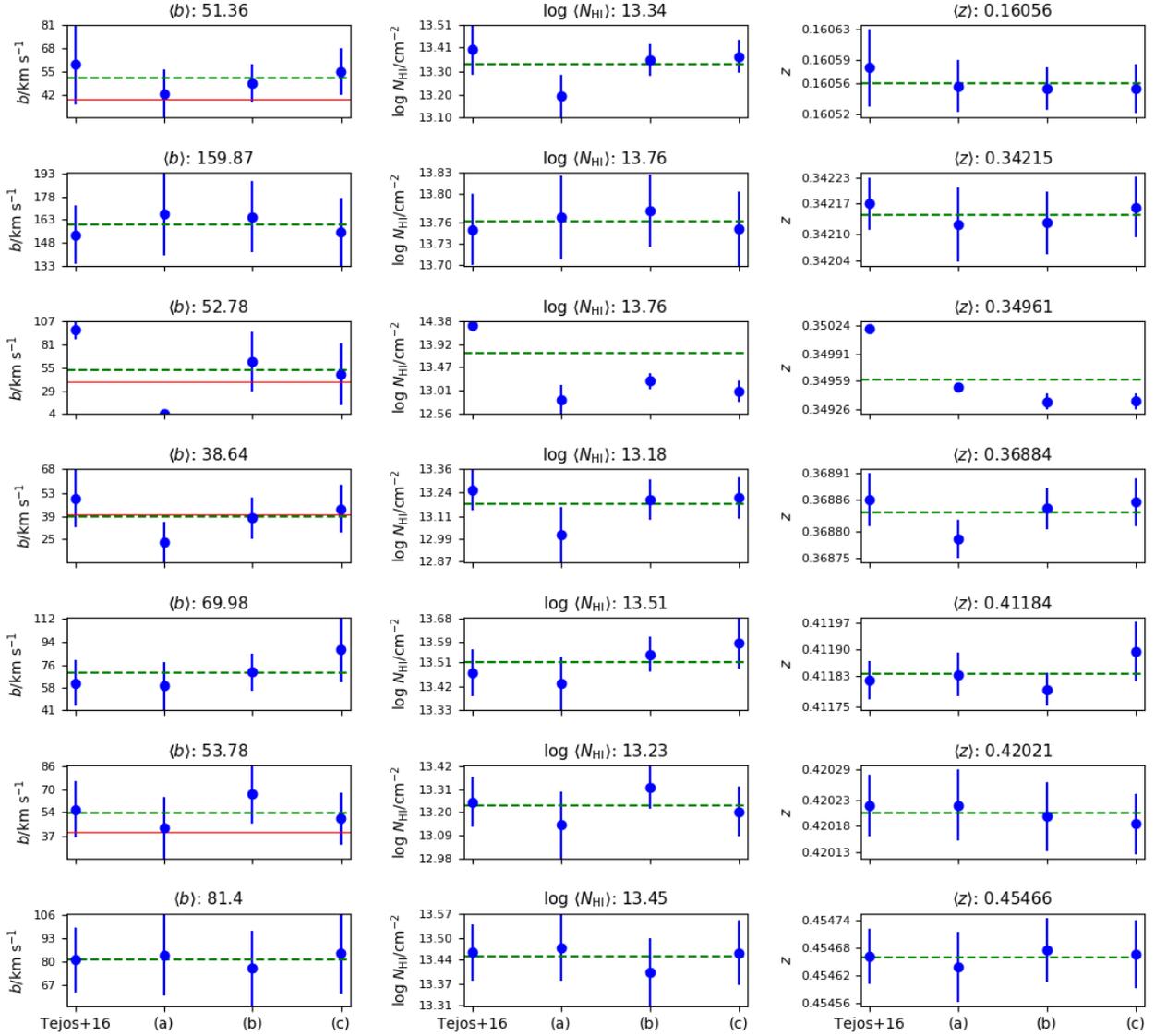}
    \caption{{Systematic comparison between the parameters obtained for the BLAs in our sample using a different data reduction, an independent continuum level estimation and  a different Voigt profile fitting software. (a), (b) and (c) represent different combination of these variables respect to those used in \citet{Tejos2016}. With the exception of the absorption feature at $z=0.3502$, all absorption features have systematic variations consistent with the level of reported statistical uncertainties. More datails in Section~\ref{sec:redo})  } }
    \label{fig:bla_comparison}
\end{figure*}

From the above, we performed the following combinations to define a set of new experiments: (a) new data and new continuum, (b) old data and new continuum, and (c) old data and old continuum, all of which performed using the new software {\sc veeper}. Figure~\ref{fig:bla_comparison} shows the obtained parameters for the putative BLAs from our different experiments and how these compare to those reported by \citet{Tejos2016}. We see that, with the exception the absorption feature at $z=0.3502$, all absorption features have systematic variations well below (or consistent) with the level of reported statistical uncertainties. The putative BLA at $z=0.3502$ is the less constrained one as it may be superimposed to a complex narrow \hi~system and its absorption profile is degenerate with that of the narrow component. Indeed, in experiment (a) this absorption feature was fit with a very narrow line (Doppler parameter $b<10$\,\kms) casting doubt on it being a genuine BLA. In any case, this feature has been also excluded from our `clean BLA sample' (for the estimation of the filament baryon fraction in Section~\ref{sec:OmegaInFil}) on the basis of the existence of a potential galaxy counterpart (see Section~\ref{sec:asso}).


\label{lastpage}
\end{document}